\PassOptionsToPackage{table,dvipsnames}{xcolor}
\documentclass[sigconf,nonacm]{acmart}

\makeatletter
\def\@ACM@checkaffil{% Only warnings
    \if@ACM@instpresent\else
    \ClassWarningNoLine{\@classname}{No institution present for an affiliation}%
    \fi
    \if@ACM@citypresent\else
    \ClassWarningNoLine{\@classname}{No city present for an affiliation}%
    \fi
    \if@ACM@countrypresent\else
        \ClassWarningNoLine{\@classname}{No country present for an affiliation}%
    \fi
}
\makeatother

\setcopyright{none}
\renewcommand\footnotetextcopyrightpermission[1]{}
\AtBeginDocument{%
  \providecommand\BibTeX{{%
    \normalfont B\kern-0.5em{\scshape i\kern-0.25em b}\kern-0.8em\TeX}}}

\usepackage[normalem]{ulem}

\usepackage{tikz}
\usepackage{amsmath}

\usepackage{filecontents}

\usepackage[hang,flushmargin]{footmisc}
\usepackage{microtype} 
\usepackage{siunitx}
\usepackage[ruled,linesnumbered]{algorithm2e}
\usepackage[utf8]{inputenc}
\usepackage{framed}
\usepackage{multirow}
\usepackage{listings}
\usepackage{verbatim}
\usepackage{tikz}
\usepackage{appendix}
\usepackage{graphicx}
\usepackage{float}
\usepackage{subfig}
\usepackage{environ}
\usepackage{comment}
\usepackage{booktabs}
\usepackage{ragged2e}
\usepackage{lipsum}
\usepackage{afterpage}
\usepackage{tabularx}
\usepackage{calc}
\usepackage{flushend}
\usepackage{bm}

\usepackage{amsbsy}

\usepackage[para]{threeparttable}

\usepackage{pifont}%
\usepackage{listings}

\definecolor{charcoal}{rgb}{0.21, 0.27, 0.31}
\definecolor{Gray}{rgb}{0.5, 0.5, 0.5}
\definecolor{slateGray}{rgb}{0.44, 0.5, 0.56}
\definecolor{smoke}{rgb}{0.52, 0.53, 0.52}
\definecolor{steelGray}{rgb}{0.44, 0.47, 0.49}

\definecolor{grey}{rgb}{0.02, 0.02, 0.01}

\newcommand{\cmark}{\textcolor{JungleGreen}{\ding{51}}}%
\newcommand{\xmark}{\textcolor{Maroon}{\ding{55}}}%
\newcommand{\blackcmark}{\textcolor{slateGray}{\ding{51}}}

\usepackage[varqu]{zi4}%
\AtBeginDocument{%
}

\definecolor{Maroon}{cmyk}{0, 0.87, 0.68, 0.32}
\definecolor{darkgreen}{rgb}{0.0, 0.2, 0.13}
\definecolor{JungleGreen}{rgb}{0.16, 0.67, 0.53}

\newcommand\addvmargin[1]{
  \node[fit=(current bounding box),inner ysep=#1,inner xsep=0]{};
}
\usetikzlibrary{fit}

\newcommand*\emptycirc[1][1ex]{%
  \begin{tikzpicture}[baseline=-4]
  \draw (0,0) circle (#1);
  \addvmargin{1mm}
  \end{tikzpicture}}%

\newcommand*\halfcirc[1][1ex]{%
  \begin{tikzpicture}[baseline=-4]
  \draw[fill] (0,0)-- (90:#1) arc (90:270:#1) -- cycle ;
  \draw (0,0) circle (#1);
  \addvmargin{1mm}
  \end{tikzpicture}}%

\newcommand*\fullcirc[1][1ex]{%
  \begin{tikzpicture}[baseline=-4]
  \fill (0,0) circle (#1);
  \addvmargin{1mm}
  \end{tikzpicture}}%

\definecolor{codegreen}{rgb}{0,0.56,0.56}
\definecolor{codegray}{rgb}{0.3,0.3,0.3}
\definecolor{codepurple}{rgb}{0.7,0,0.82}
\definecolor{codeblack}{rgb}{0,0,0}
\definecolor{backcolor}{rgb}{0.99,0.99,0.99}
\definecolor{keywordblue}{rgb}{0.1,0.1,1}
\definecolor{commentgray}{rgb}{0.65,0.65,0.65}

\lstdefinestyle{mystyle}{
    frame=single,
    backgroundcolor=\color{backcolor},   
    commentstyle=\sffamily\it\color{commentgray},
    keywordstyle=\color{keywordblue},
    numberstyle=\tiny\color{codeblack},
    stringstyle=\color{codepurple},
    basicstyle=\rmfamily\scriptsize\color{black},
    breakatwhitespace=false,         
    breaklines=true,                 
    captionpos=b,                    
    keepspaces=true,                 
    numbers=left,                    
    numbersep=4pt,                  
    showspaces=false,                
    showstringspaces=false,
    showtabs=false,                  
    tabsize=4,
    float=tp,
    floatplacement=tbp,
    abovecaptionskip=0pt
}

\newcommand{\name}{\textsc{PrefetchX}}

\definecolor{observationColor}{rgb}{0.9,0.97,0.95} 
\newcommand{\evictObs}{\textbf{\textit{Observation~1}}}
\newcommand{\flushObs}{\textbf{\textit{Observation~2}}}

\definecolor{usecaseColor}{rgb}{0.98,0.94,0.97} 
\newcommand{\flushscaUsecase}{\textbf{\textit{Use-Case~2}}}
\newcommand{\evictscaUsecase}{\textbf{\textit{Use-Case~1}}}

\begin{document}

\title{New Cross-Core Cache-Agnostic and Prefetcher-based Side-Channels and Covert-Channels}

\author{Yun Chen}
\affiliation{%
  \institution{National University of Singapore}
}

\author{Ali Hajiabadi}
\affiliation{%
  \institution{National University of Singapore}
}

\author{Lingfeng Pei}
\affiliation{%
  \institution{National University of Singapore}
}

\author{Trevor E. Carlson}
\affiliation{%
  \institution{National University of Singapore}
}

\begin{abstract}
In this paper, we reveal the existence of a new class of prefetcher, the XPT prefetcher, in the modern Intel processors which has never been officially documented. It speculatively issues a load, bypassing last-level cache (LLC) lookups, when it predicts that a load request will result in an LLC miss. We demonstrate that XPT prefetcher is shared among different cores, which enables an attacker to build cross-core side-channel and covert-channel attacks.
We propose \name{}, a cross-core attack mechanism, to leak users' sensitive data and activities. 

We empirically demonstrate that \name{} can be used to extract private keys of real-world RSA applications. Furthermore, we show that \name{} can enable side-channel attacks that can monitor keystrokes and network traffic patterns of users. Our two cross-core covert-channel attacks also see a low error rate and a 1.7MB/s maximum channel capacity. 
Due to the cache-independent feature of \name{}, current cache-based mitigations are not effective against our attacks. Overall, our work uncovers a significant vulnerability in the XPT prefetcher, which can be exploited to compromise the confidentiality of sensitive information in both crypto and non-crypto-related applications among processor cores. 
\end{abstract}

\maketitle

\pagestyle{plain}

\section{Introduction}

Decades of research in designing efficient and modern processors has resulted in various performance enhancements at the microarchitectural level, like out-of-order execution, speculative execution, multi-core processing, caching, prefetching, and sharing resources among different cores. However, in recent years there has been rapid discovery of security vulnerabilities arising from these performance enhancement techniques~\cite{Kocher2018spectre, chen2023afterimage, puddu2020frontal, gast2022squip, deng2022leaky, yarom2014flush+, disselkoen2017prime+, gruss2016flush+, lipp2018meltdown,xiao2023hacky}.
These vulnerabilities are exploited to either infer 
a user's
private data and secret keys 
(in case of side-channel attacks) or to stealthily transfer data in the system (in the case of covert-channel attacks).

One of the prominent sources of information leakage in modern processors are prefetchers since they can exhibit a footprint of the data accessed by the victims, similar to caches.
The goal of prefetching is to bring the data closer to the core if it has high confidence to be needed in the near future. This technique can be implemented on either hardware~\cite{intel_prefetcher, ipstride} or software~\cite{GUTTMAN2015401}.
Recent attacks exploit a variety of %
prefetching mechanisms~\cite{chen2023afterimage, guo2022adversarial, guo2022leaky, host, ccs, gruss2016prefetch, fang2018prefetch, vicarte2022augury} to leave persistent malicious and secret-dependent changes in the system that can be later inferred by the attacker.
In this paper, we explore a prefetcher in Intel processors, called eXtended Prediction Table (XPT), that is located in parallel to the LLC. The XPT prefetcher is not documented in the latest Intel official architecture manual~\cite{intel2019} but only has a brief functionality description in the HPC optimization manual~\cite{xpt}.

In this paper, we first reverse-engineer the XPT prefetcher and reveal its prefetching mechanism. We investigate the interaction of different cores through the prefetcher. %
Our analysis has resulted in %
the first side-channel using the XPT prefetcher, called \name{}, which can be exploited in $3^{rd}$ generation Xeon processors\footnote{The latest Xeon processors at the time of paper submission.} to effectively monitor the victim's page activities. %
This is achieved by deliberately mistraining the XPT prefetcher on specific pages and subsequently examining the prefetcher's status. Consequently, the attacker can reconstruct the victim's sensitive information, like cryptographic keys.

Unlike many standard threat models that necessitate the victim and attacker to share the same physical core~\cite{Kocher2018spectre, chen2023afterimage, host, puddu2020frontal, gast2022squip, ccs, zhao2022binoculars, vicarte2022augury, deng2022leaky}, \name{} enables \textit{cross-core} 
side-channel attacks. This substantially broadens the scope of potential targets and amplifies the potential security impact.
In addition, the \name{} attacks do not rely on the cache subsystem; 
we do not rely on the caches %
as a source of leakage nor as a primitive to check the prefetcher's status.
To the best of our knowledge, we are the first to identify, explore, and reverse-engineer the XPT prefetcher. %
This discovery highlights the existence of hidden channels that could expose security risks in processor architectures.

Our key contributions in this work are as follows:
\begin{itemize}
    \item 
    We uncover a briefly-mentioned %
    prefetcher, named the XPT prefetcher, in the $3^{rd}$ generation of Intel server processors.
    We fully reverse-engineer this %
    prefetcher, and provide a detailed characterization of its features and behaviors (Section~\ref{sec:reverse-eng}).

    \item We construct four end-to-end cross-core side-channel attacks using the XPT prefetcher as the leaky source. These attacks include a keystroke attack, a network traffic monitoring attack, and two distinct RSA attacks to break the private exponent (a square-and-multiply RSA used in GnuPG 1.4~\cite{gnupg} and a timing-constant Montgomery-Ladder RSA~\cite{rsaml} used in OpenSSL~\cite{opensslrsaml} and MbedTLS~\cite{mbed}). Our results demonstrate the practicality and effectiveness of \name{} in real-world scenarios\footnote{We have responsibly disclosed our findings to the Intel PSIRT team and have received approval to distribute these details.} (Section~\ref{sec:eviction-sca} and Section~\ref{sec:tlb_sca}).

    \item We develop two cross-core covert-channel attacks with low error rates, further demonstrating the high applicability and potential security implications of \name{} (Section~\ref{sec:covert-channel-attacks}). 
\end{itemize}

\section{\name{} Motivation and Overview}
\label{sec:overview}

\subsection{The XPT Prefetcher as a Leakage Source}
\label{sec:xpt-overview}

\begin{figure}[t]
    \centering
    \includegraphics[trim=0 0cm 0 0cm,width=1\linewidth]{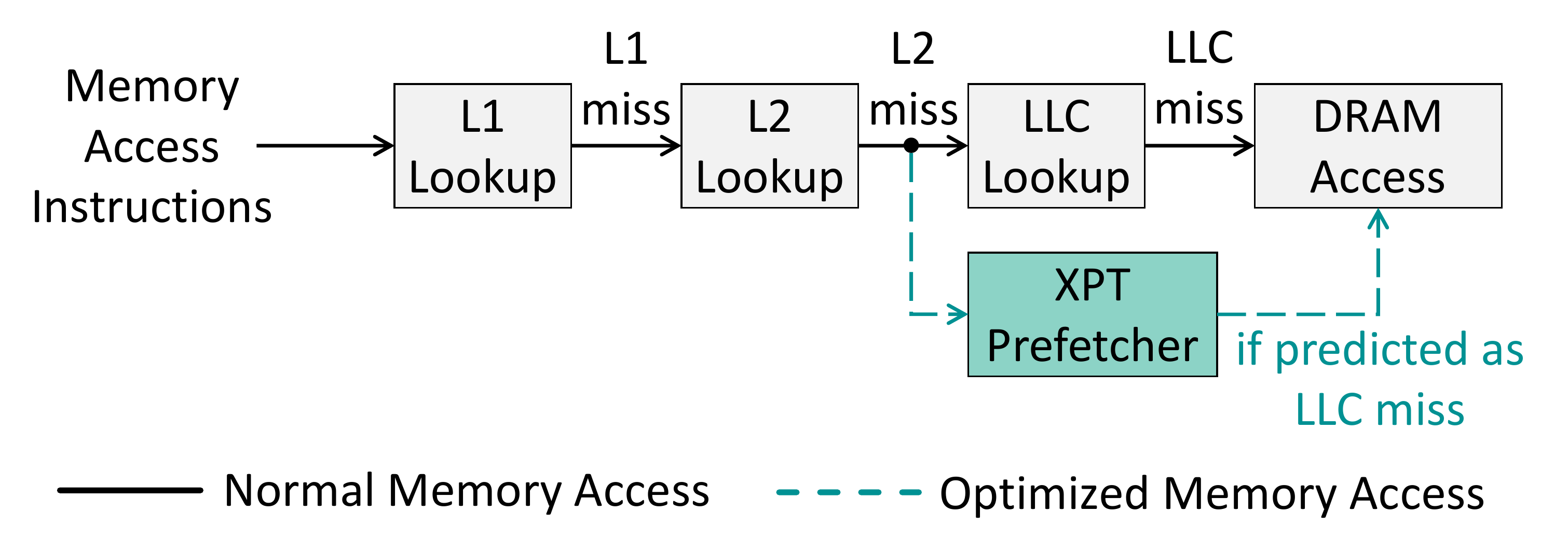}
    \caption{Overview of the XPT prefetcher operation.}
    \label{fig:xpt-overview}
\end{figure}

\name{} exploits the XPT prefetcher in Intel processors, and the root cause of the attack is the special mechanism of this component.
The XPT prefetcher resides in parallel with the LLC
and its main goal is to predict LLC misses and to send speculative requests to the DRAM to conduct LLC lookups in advance. This can reduce memory access latency in the case of a correct prediction.
Hence, the memory accesses sent to the LLC subsystem exhibit three levels of access latency (we use the \texttt{RDTSC} to compute the latency): 
(1) an \textit{LLC hit} (less than 160 cycles in our setup and experiments), (2) \textit{LLC miss} (350+ cycles), and finally (3) \textit{optimized LLC miss} (170-330 cycles). The optimized LLC misses occur in cases that the XPT prefetcher has a correct prediction and is enabled. 
Figure~\ref{fig:xpt-overview} shows an overview of the XPT prefetcher alongside the cache hierarchy. 
This timing variation is tightly coupled with the memory activities of the executing program that enables easy status monitoring of the prefetcher.

In addition, our detailed experiments in Section~\ref{sec:reverse-eng} reveal that (1) XPT prefetching can be trained and triggered \textit{across different cores}, that (2) the LLC miss prediction is based on a miss counter in the prefetcher for a 4\,KiB page granularity, and that (3) different entries in the prefetcher are indexed based on the physical address of the data and distinct by Address Space ID (ASID) and Thread ID (TID)\footnote{Note, each process has a unique ASID, whereas different threads in the same process share an ASID but are assigned different TIDs}.

In this work, we make two key observations from the XPT prefetcher behavior that enables us to build side-channel and covert-channel attacks. First, \textbf{\textit{we observe that the XPT prefetcher entries are evicted based on the Least Recently Used (LRU) policy if an unmapped page in the same ASID is accessed}}. In other words, if thread \texttt{t1} trains the prefetcher with a certain set of data pages and thread \texttt{t2} accesses a page outside this set then the oldest page trained by \texttt{t1} is evicted from the XPT prefetcher. 
Second, \textbf{\textit{we observe that flushing the TLB resets the XPT prefetcher status}}.
We demonstrate that if process 1 trains the XPT prefetcher with a shared page A and process 2 flushes that page mapping from TLB then the XPT prefetcher will not be triggered for the previously trained page A.
In Section~\ref{sec:root-causes}, we provide the details of our observations.

\subsection{\name{} Threat Model and Attack Surface}
We assume two unprivileged threads/processes running on two different cores of the same processor, corresponding to the attacker and victim in the side-channel attacks, or the sender and receiver in the covert-channel attack.  %
We aim to transfer or leak private information \textit{across cores inside a processor}\footnote{The interplay with Intel Software Guard Extension (SGX) 
is considered to be outside the scope of this work as SGX was not designed to be secure against side-channel attacks~\cite{sgx_sca} and is not supported in AWS (our test platform).}
by observing the page-accessing activities through the XPT prefetcher. %
We also assume that the two parties have shared data such as shared memory space or shared libraries (\textit{e.g.} OpenSSL~\cite{OpenSSL}), similar to prevalent hardware attacking scenarios~\cite{advprefetch, gruss2016flush+, vicarte2022augury, yarom2014flush+}. As we leverage \texttt{RDTSC} to compute the access latency, which is running with a constant frequency and does not change with the CPU frequency~\cite{intel_software}, we make no assumptions on the CPU frequency of targeted cores.

\subsection{\name{} Workflow}
The general attack flow of \name{} can be summarized as three main steps:

\textbf{Step 1: Channel Preparation}. To construct a side-channel, the attacker first identifies the physical address of the target shared page that the victim's interesting actions will access, which can be easily done by using established techniques~\cite{kwong2020rambleed, zhao2022binoculars}. %
Then the attacker primes a targeted XPT entry corresponding to the victim's physical address on the shared page. Based on the findings of our reverse-engineering analysis (refer to Section~\ref{sec:trigger_xpt}), generating 32 cache misses on a page is sufficient to prime the XPT prefetcher to begin prefetching. %
When using the XPT prefetcher as a covert-channel, the sender and the receiver leverage a predefined page for communication.

\textbf{Step 2: Performing the target action}. After the side-channel setup, the victim execution is initiated. As the victim runs, the prefetcher's status is evicted or reset if the target action is invoked.
In the case of covert-channel, the sender either primes or flushes/evicts an XPT entry to send different bits of the secret message (\textit{i.e.}, b'0 or b'1).

\textbf{Step 3: Check the XPT prefetcher status}. The attacker checks the prefetcher status to determine whether the victim has executed the target function (in case of covert-channel, to check what bit has been transmitted by the sender). 

By following these steps, \name{} is able to effectively leak the victim's sensitive page-related actions through side-channel analysis, while also %
enabling a covert channel. %
Section~\ref{sec:eviction-sca} and Section~\ref{sec:tlb_sca} describe our side-channel attacks, while Section~\ref{sec:covert-channel-attacks} describes the details of how to use \name{} as a 
covert-channel.

\section{Background}
\label{sec:background}

\subsection{Prefetching}
The main goal of a prefetcher is to predict the data and instructions that will be required by the processor in the near future and to fetch them from the main memory into the cache in advance, improving system performance. There exist a variety of prefetching styles. Software prefetching allows programmers to insert prefetching instructions in desired locations of the code~\cite{gruss2016prefetch, guo2022leaky, advprefetch, intel_software}. Hardware prefetchers, such as next-line prefetcher~\cite{intel_prefetcher}, stride prefetcher~\cite{chen2023afterimage}, spatial and temporal prefetchers~\cite{somogyi2006spatial, 9229804, 8327004, 7820158}, each have different strategies for predicting and fetching data and can be beneficial based on the application running in the hardware.

\begin{table}[tb]
\centering
\scalebox{0.75}{
\begin{tabular}{c|c|c|c}
\toprule
Intel Prefetcher                                                           & Location & Pattern  & Attacks                                                                        \\ \midrule
\begin{tabular}[c]{@{}c@{}}Data Cache Unit\\ (DCU)\end{tabular}            & L1-D     & Next cache line~\cite{intel_prefetcher} & -                                                                 \\ \hline
\begin{tabular}[c]{@{}c@{}}Instruction pointer\\ (IP)-stride\end{tabular} & L1-D      & \begin{tabular}[c]{@{}c@{}}Load instructions with \\ regular stride~\cite{chen2023afterimage}\end{tabular} & \begin{tabular}[c]{@{}c@{}}~\cite{chen2023afterimage,host} \\ ~\cite{ccs} \end{tabular}\\ \hline
\begin{tabular}[c]{@{}c@{}}Data prefetch logic\\ (DPL)\end{tabular}        & L2       & \begin{tabular}[c]{@{}c@{}}128-bytes-aligned~\cite{intel_prefetcher} \\ pair cache line\end{tabular}  & - \\ \hline
Streamer                                                                   & L2       &  
\begin{tabular}[c]{@{}c@{}}Several cache lines  \\ backward or forward~\cite{9229804} \end{tabular}  & ~\cite{9229804}                                                     \\ \hline
\rowcolor{JungleGreen!10}%
\textcolor{black}{\textbf{XPT prefetcher}} & \textcolor{black}{\textbf{LLC}} & \textcolor{black}{\textbf{LLC miss predictor}} & \textcolor{black}{\textbf{This work}}\\ 
\bottomrule
\end{tabular}}
\caption{Intel hardware prefetchers~\cite{whitepaper}. %
}
\label{tb:prefs}
\end{table}

According to Intel's whitepaper~\cite{whitepaper}, their processor designs feature four hardware prefetchers, which are listed in Table~\ref{tb:prefs}. 
In this work, we focus on the eXtension Prediction Table (XPT) prefetcher, which is not documented by Intel in any of their hardware reference manuals~\cite{intel2019, intel_software}, but instead was briefly described %
in the HPC optimization reference manual~\cite{xpt} for the $3^{rd}$ generation of Xeon processors. %
The key difference of the XPT prefetcher with others is that it is placed at the last-level cache (LLC) (see Table~\ref{tb:prefs}) which makes it a potential channel for cross-core attacks.

\subsection{Translation Lookaside Buffer and Page Tables}
\label{sec:tlb-background}

\begin{figure}[t]
    \centering
    \includegraphics[trim=0 0.5cm 0 0cm,width=1\linewidth]{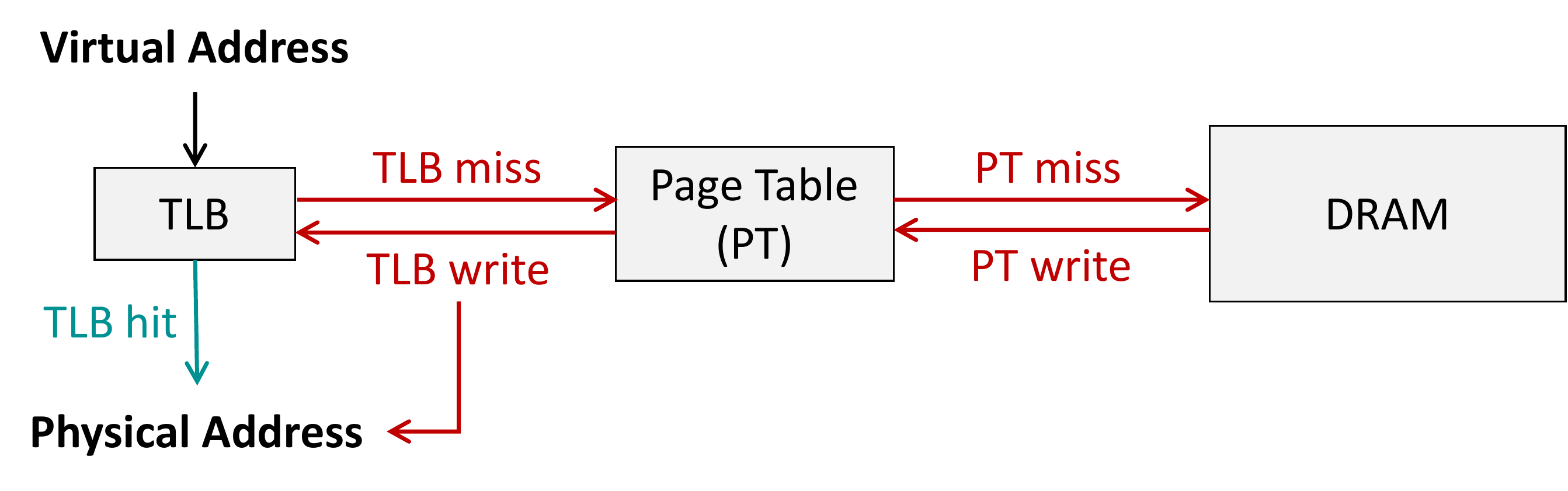}
    \caption{Process of a page walking.}
    \label{fig:memtran}
\end{figure}

\textbf{Translation and page walking}.
Figure~\ref{fig:memtran} shows the virtual address to physical address translation process in x86 processors. %
The memory management unit (MMU) has a cache, called translation lookaside buffer (TLB), that keeps the most recently accessed page mappings for a fast translation in case of a TLB hit. In case of a TLB miss, the MMU performs a page walk over the page tables (PT) %
to find the mapping. The operating system (OS) is responsible for maintaining the mappings from the virtual addresses provided by each process to the physical addresses in dynamic random-access memory (DRAM).  %
As these mappings typically have a granularity of 4KiB, the page walking process involves primarily the upper 48 bits of the virtual address, referred to as the \textit{page offset}. Meanwhile, the lower 12 bits remain identical to those of the physical address and are called the \textit{in-page offset}.

\textbf{TLB flush.} In modern processors, the TLB is usually flushed at any time if the CPU needs to update the TLB~\cite{intel_software}. However, there are three main events that require a TLB flush:

\begin{itemize}
    \item\textbf{Context Switch:} When the OS switches from one process to another, the TLB needs to be updated with the new virtual-to-physical address mappings for the new process\footnote{However, frequent TLB flushes during context switches are avoided in modern processors as every TLB entry is tagged with an ASID, ensuring minimal impact on process performance.}. 
    \item \textbf{Page Table Changes:} If the page tables are modified, such as when a page is swapped in or out of memory, the TLB may need to be flushed to ensure the correct page mappings are used.
    \item \textbf{Inter-Processor Interrupt for Cache Coherency:} In a multiprocessor system, an inter-processor interrupt (IPI) is a type of interrupt used to schedule tasks or synchronize data between different processors or cores. When multiple threads run on different cores and share memory, maintaining cache coherency between the cores is critical. In some cases, when one thread updates shared memory, the operating system (OS) may launch an IPI when the thread is switched to another core, followed by flushing the TLB to ensure that the correct data is presented.
\end{itemize}

We will later show that TLB flush impacts the XPT prefetcher status and is the root cause of a number of our cross-core covert-channel and side-channel attacks.

\section{Reverse-Engineering the XPT Prefetcher}
\label{sec:reverse-eng}
To %
better understand %
the XPT prefetcher, we performed extensive microbenchmarking to reverse engineer it. In this paper, we use an AWS EC2 m6i.4xlarge instance powered by a 16-core Intel(R) Xeon(R) Platinum 8375C CPU (Ice Lake generation, Sunny Cove microarchitecture).

\subsection{Triggering the XPT Prefetcher}
\label{sec:trigger_xpt}
Based on our experiments, the XPT prefetcher is enabled after a fixed number of LLC cache misses.
To determine the number of LLC cache misses required to trigger the XPT prefetcher, we use a microbenchmark outlined in Listing~\ref{listing:xpt_trigger}. The microbenchmark first initializes a page (line 20) and flushes the initialized data from the cache using \texttt{clflush} instructions (line 23). 
We then train the XPT prefetcher %
up to a specific
number of LLC misses, and finally, test the memory access latency for an LLC cache miss (\textit{i.e.}, DRAM access) to test if XPT is enabled.
The parameter \texttt{n} in the \texttt{train()} function specifies the number of cache misses to be generated, which is achieved through the use of the \texttt{generate\_random()} function that generates \texttt{n} unique random numbers as indices for accessing the page. This ensures that the memory accesses are irregular and will not trigger other potential prefetchers, such as the next-line, IP-stride, adjacent, and streamer prefetchers. After generating \texttt{n} cache misses, a \texttt{test\_index} is set, computed at runtime and distant from \texttt{n}, to test the DRAM latency of the access at \texttt{ptr[test\_index]}.

\begin{figure}[t]
\begin{lstlisting}[language=C, caption=Determining how to trigger the XPT prefetcher. The results shown in Figure~\ref{fig:trigger_xpt}., label=listing:xpt_trigger]
void train(char *ptr) {
    int n = 32;
    int random_num[n];
    generate_random(random_num);
    for (int i = 0; i < n; i++) {
        int index = random_num[i];
        char junk = ptr[index * CACHE_LINE];
    }
}
void test(char *ptr) {
    int test_index = rand() % 10 + 53;
    int start = rdtsc();
    asm volatile("mfence");
    char junk = ptr[test_index * CACHE_LINE];
    asm volatile("lfence");
    int diff = rdtsc() - start;
}
void main() {
    int page_size = 4096;
    char *ptr = (char *)mmap(page_size, ..., MAP_LOCK, ...);
    for (int i = 0; i < page_size; i+=64)
      ptr[i] = 'x';
    flushall(ptr);
    train(ptr);
    test(ptr);
}
\end{lstlisting}
\end{figure}

\begin{figure}[t]
    \centering
    \includegraphics[trim=0 0.3cm 0 1cm,width=\linewidth]{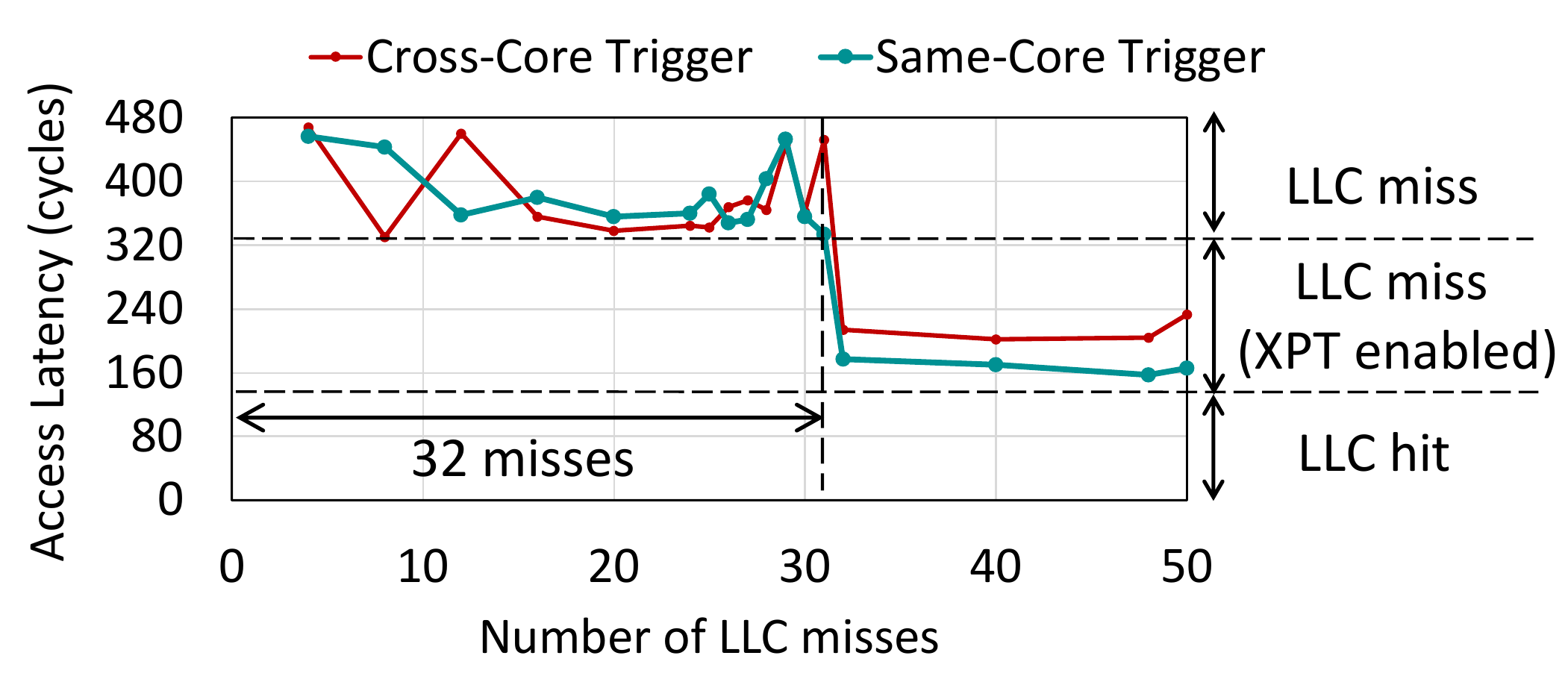}
    \caption{Triggering the XPT prefetcher from either the same core or a different core. 
    In both cases, the XPT prefetcher is triggered after 32 LLC misses. %
    }
    \label{fig:trigger_xpt}
\end{figure}

The results of our experiments are presented in Figure~\ref{fig:trigger_xpt} (green line) which runs our microbenchmark with %
an increasing number
of LLC misses during %
training.
It is evident that 
One can see that the XPT prefetcher begins to start requesting data early
after 32 cache misses, leading to a reduction in DRAM access latency from approximately 350+ cycles to an optimized LLC cache miss (between 160 to 300 cycles) which has triggered
the XPT prefetcher. 

As the XPT is located in parallel to the LLC, it is reasonable to assume that it is shared between two different cores. To verify this assumption, we modify the Listing~\ref{listing:xpt_trigger} to run the \texttt{train} and \texttt{test} functions on different cores. The result is presented in Figure~\ref{fig:trigger_xpt} (red line), which is similar to our observation from triggering the XPT from the same core
(green line). This implies that \textit{the XPT prefetcher is shared between different cores.}

\subsection{Indexing into the XPT Prefetcher}
Intel commonly employs the instruction pointer (IP)~\cite{chen2023afterimage}, virtual address~\cite{intel2019}, or physical address~\cite{disselkoen2017prime+} as an index %
for accessing hardware tables or caches in its processors. %
Our objective is to examine and understand the indexing mechanism employed by the XPT prefetcher. Given its position as a bypass for LLC lookup and its location between the LLC and L2 cache, our initial focus is to determine if the XPT utilizes physical address indexing, similar to the LLC~\cite{disselkoen2017prime+,intel2019}.

\begin{figure}[t]
\begin{lstlisting}[language=C, caption=Determining how the XPT prefetcher table is indexed. We use semaphore to control the train/test sequence, label=listing:index]
void main() {
    int page_size = 4096;
    char *ptr = (char *)mmap(page_size, ..., MAP_LOCK, ...);
    pid_t sub_process = fork();
    if (sub_process > 0) // main process
        train(ptr);
    else if (sub_process == 0) // sub-process
        test(ptr);
}
\end{lstlisting}
\end{figure}

\begin{table}[]
\centering
\scalebox{0.85}{
\begin{tabular}{c||c|c}
\toprule
Tested Scenario & Indexing Policy & Is XPT triggered? \\
\midrule
Scenario-1 & Physical Page &  \cmark \\ 
\hline
Scenario-2 &  Virtual Page &\textcolor{Maroon}{\xmark} \\
\hline
Scenario-3 & Instruction Pointer  &  \textcolor{Maroon}{\xmark} \\  
\bottomrule
\end{tabular}}
\caption{The XPT prefetcher triggering results with different indexing policies.}
\label{tb:index}
\end{table}

We developed a microbenchmark (See Listing~\ref{listing:index}) to evaluate the indexing policy of the XPT prefetcher. Our microbenchmark involves allocating a shared memory page and a child process. The \texttt{train} and \texttt{test} functions, as shown in Listing~\ref{listing:xpt_trigger}, are reused in this experiment. The main process trains the XPT prefetcher on the shared page and the child process accesses an uncached data block in the shared page to determine if the XPT prefetcher has been triggered.

A shared page has a unique mapping in the virtual address spaces of both the main process and the child process (in Listing~\ref{listing:index}), because the main process and child process have distinct address spaces. However, the physical page mapping is the same for both processes. The results of our experiment, as shown in the first row of Table~\ref{tb:index} (Scenario-1), indicate that the XPT prefetcher is consistently triggered when two processes access the same physical page, suggesting that \textit{the XPT is indexed by the physical page}. We also tested the possibility of indexing by the virtual address and IP (Scenario-2 and Scenario-3 Table~\ref{tb:index}, respectively), but the results indicate that these methods do not trigger the XPT prefetcher and are not used for indexing the XPT prefetcher.

To gain a deeper understanding of how many bits of the physical address are utilized for indexing the XPT, we create a 32\,GiB memory pool so that we can manipulate the least significant 35 bits of physical addresses. We then identify two physical pages with identical least significant $M$ bits for their page offsets (\textit{i.e.}, physical addresses from $13^{th}$ bit to the $(13 + M)^{th}$ bit are the same). By training the XPT prefetcher on one page, we expect that if the table is indexed by the lower $M$ bits of the page, another page should also trigger the prefetcher. Nevertheless, we do not observe this behavior even when $M$ is increased to 20, suggesting that 
\textit{the XPT prefetcher should be either indexed by the full physical address or possesses a tag for each physical page.}

\subsection{Page Boundary}

\begin{figure}[t]
\begin{lstlisting}[language=C, caption=Determine the page boundary of the XPT prefetcher., label=listing:page_boundary]
void main() {
    char *huge_1GiB_page = (char *)mmap(1 << 30,
                    ..., MAP_HUGE_1GB|MAP_LOCK);
    char *huge_2MiB_page = (char *)mmap(1 << 21, 
                    ..., MAP_HUGE_2MB|MAP_LOCK);
    char *normal_4KiB_page = (char *)mmap(8192, 
                                 ..., MAP_LOCK);
    char *ptr; //refers to one of pools above
    train(ptr);
    test(ptr[4096]); //cross 4KiB boundary
}
\end{lstlisting}
\end{figure}

\begin{table}[t]
\centering
\scalebox{0.75}{
\begin{tabular}{c||ccc}
\toprule
\multirow{3}{*}{Page Table Size}& \multicolumn{3}{c}{Cross 4KiB-Page Boundary Trigger} \\ 
\cline{2-4} &
\multicolumn{1}{c|}{\begin{tabular}[c]{@{}c@{}}Contiguous on\\ Virtual Address\end{tabular}} &
\multicolumn{1}{c|}{\begin{tabular}[c]{@{}c@{}}Contiguous on\\ Physical Address\end{tabular}} &
\multicolumn{1}{c}{\begin{tabular}[c]{@{}c@{}}Is XPT\\ Triggered?\end{tabular}} \\ 
\midrule
1\,GiB & \multicolumn{1}{c|}{Yes} & \multicolumn{1}{c|}{Yes} & \xmark  \\ \hline
2\,MiB & \multicolumn{1}{c|}{Yes} & \multicolumn{1}{c|}{Yes} & \xmark  \\ \hline
4\,KiB & \multicolumn{1}{c|}{Yes} & \multicolumn{1}{c|}{No} & \xmark \\ 
\bottomrule
\end{tabular}}
\caption{The XPT prefetcher triggering results with various page tables setting.}
\label{tb:page_boundary}
\end{table}

\begin{figure}[t]
\begin{lstlisting}[language=C, caption=Determine the number of entries of the XPT prefetcher., label=listing:xpt_entry]
void train(char *ptr, int num) {
    int n = 32;
    int random_num[n];
    generate_random(random_num);
    for (int i = 0; i < num; i++) {
      for (int j = 0; j < n; i++) {
          int idx = random_num[j];
          char junk = ptr[i*4096+idx*CACHE_LINE];
      }
    }
}
\end{lstlisting}
\end{figure}

\begin{figure}[t]
    \centering
    \includegraphics[trim= 0 4.4cm 0 1cm,width=1\linewidth]{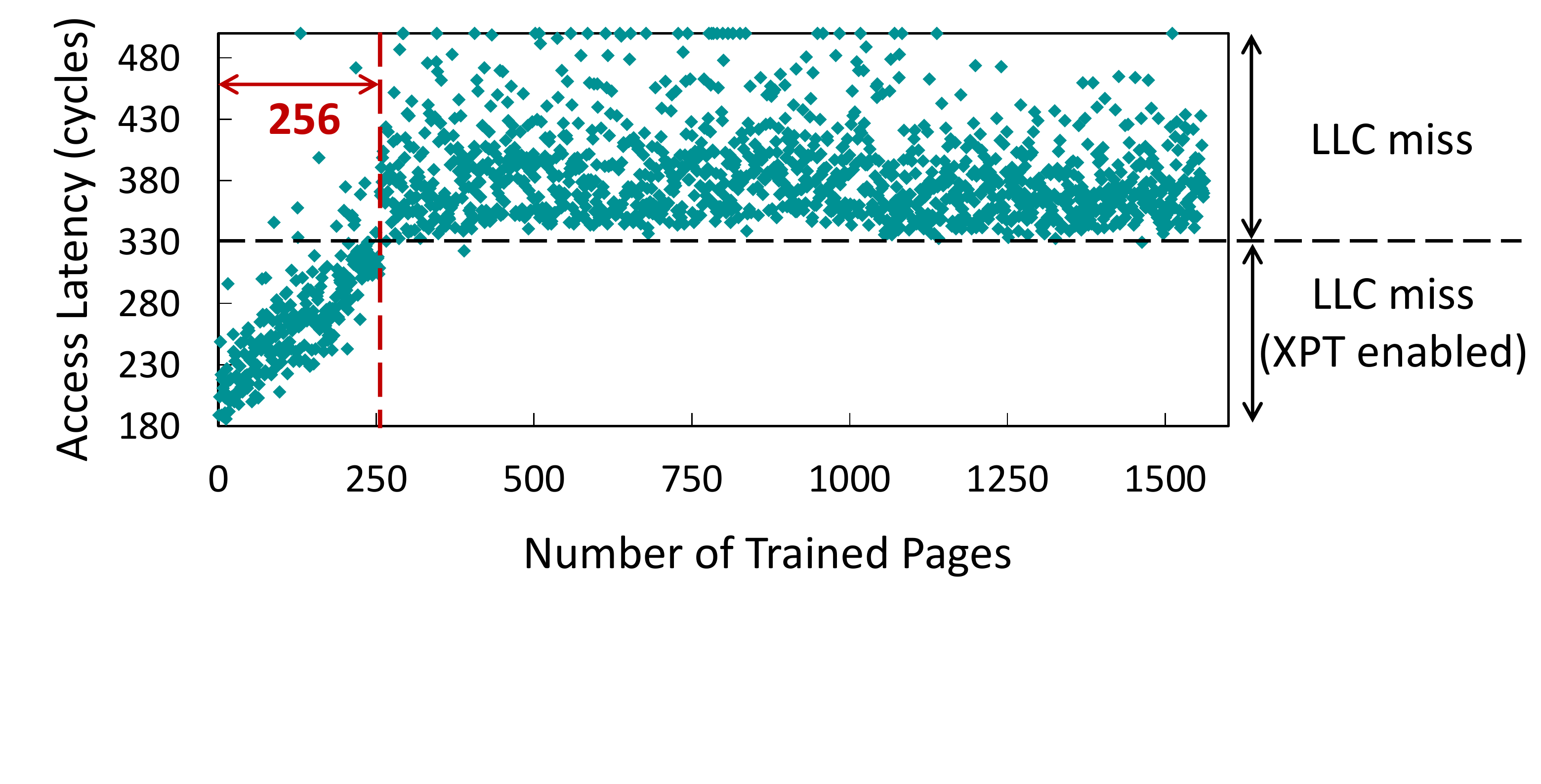}
    \caption{The number of entries of the XPT prefetcher.}
    \label{fig:entry_xpt}
\end{figure}

To determine if the XPT prefetcher is enabled in the case of huge-paging (larger than normal size of 4\,KB) and cross-4\,KB accessing.
The benchmark shown in the Listing~\ref{listing:page_boundary} first maps three memory pools named (1) \texttt{huge\_1GiB\_page}, (2) \texttt{huge\_2MiB\_page}, and (3) \texttt{normal\_4KiB\_page}. The first two memory pools will be allocated with huge page tables (1\,GiB or 2\,MiB), and the last one will be allocated with a normal page table, \textit{i.e.}, 4\,KiB. All these pages will be labeled with \texttt{MAP\_LOCK} to avoid unused pools being reclaimed and impacting the experimental results. %
We then train the XPT prefetcher on the 4\,KiB region and test it on another 4\,KiB region. In this case, only \texttt{normal\_4KiB\_page} has the cross page accesses, and the XPT prefetcher should be re-trained on the new page. The result presented in Table~\ref{tb:page_boundary}, however, shows that even \texttt{huge\_1GiB\_page} and \texttt{huge\_2MiB\_page} cannot trigger the XPT prefetcher after crossing the 4\,KiB boundary, which indicates that \textit{the XPT prefetcher has a 4\,KiB boundary tag in every entry that prevents cross 4\,KiB-page prefetch}.

\subsection{Number of Entries and Set Associativity}
\label{sec:entry}
We construct a microbenchmark that trains the XPT prefetcher on varying numbers of 4KiB pages (Listing~\ref{listing:xpt_entry}). By giving different numbers to \texttt{num}, we can create \texttt{num} entries in the XPT prefetcher. After training, we then use the \texttt{test} function shown in Listing~\ref{listing:xpt_trigger} to test if the first trained page can still trigger the XPT prefetcher.

\begin{table*}[t]
\centering
\scalebox{0.95}{
\begin{threeparttable}[b]
\begin{tabular}{c||c|c|c|c|c|c|c}
\toprule
  Scenario &
  \begin{tabular}[c]{@{}c@{}}Cross \\ Process\end{tabular} &
  \begin{tabular}[c]{@{}c@{}}Cross \\ Thread\end{tabular} &
  \begin{tabular}[c]{@{}c@{}}Cross \\ Core$^*$\end{tabular} &
  \begin{tabular}[c]{@{}c@{}}Same \\ Core\end{tabular} &
  Training Operation &
  Invalidating Operation &
  Invalidate? \\ 
\midrule
Scenario-1 & \blackcmark
   & \blackcmark 
   & \blackcmark
   & \blackcmark
   &
  \begin{tabular}[c]{@{}c@{}}Process 1, page A\end{tabular} &
  \begin{tabular}[c]{@{}c@{}}Process 2, page A\end{tabular} &
  \begin{tabular}[c]{@{}c@{}}Process 1, page A, \cmark  \textbf{\textcolor{JungleGreen}{trigger}}\end{tabular}
    \\ \hline
Scenario-2 & \blackcmark
   & \blackcmark
   & \blackcmark
   & \blackcmark
   & 
  \begin{tabular}[c]{@{}c@{}}Process 1, page A\end{tabular} &
  \begin{tabular}[c]{@{}c@{}}Process 2, page B\end{tabular} &
  \begin{tabular}[c]{@{}c@{}}Process 1, page A,B, \cmark  \textbf{\textcolor{JungleGreen}{trigger}}\end{tabular} \\ \hline
Scenario-3 & -
   & \blackcmark 
   & \blackcmark
   & \blackcmark
   &
  \begin{tabular}[c]{@{}c@{}}Thread 1, page A\end{tabular} &
  \begin{tabular}[c]{@{}c@{}}Thread 2, page A\end{tabular} &
  \begin{tabular}[c]{@{}c@{}}Thread 1, page A, \cmark  \textbf{\textcolor{JungleGreen}{trigger}}\end{tabular} \\ \hline
Scenario-4 & -
   & \blackcmark
   & \blackcmark
   & \blackcmark
   &
  \begin{tabular}[c]{@{}c@{}}Thread 1, page A\end{tabular} &
  \begin{tabular}[c]{@{}c@{}}Thread 2, page B\end{tabular} &
  \begin{tabular}[c]{@{}c@{}}Thread 1, page A, \xmark \textbf{\textcolor{Maroon}{\,no trigger}}\end{tabular} \\ \hline
Scenario-5 & -
   & \blackcmark
   & \blackcmark
   & \blackcmark
   & 
  \begin{tabular}[c]{@{}c@{}}Thread 1, pages A, B\end{tabular} &
  \begin{tabular}[c]{@{}c@{}}Thread 2, page C\end{tabular} &
  \begin{tabular}[c]{@{}c@{}}\begin{tabular}[l]{@{}c@{}}Thread 1, page A, \xmark \textbf{\textcolor{Maroon}{\,no trigger}}\end{tabular}\\ \begin{tabular}[l]{@{}c@{}}Thread 1, page B, \cmark  \textbf{\textcolor{JungleGreen}{trigger}}\end{tabular}\end{tabular} \\ \hline
Scenario-6 & \blackcmark
   & \blackcmark
   & \blackcmark
   & \blackcmark
   & 
  \begin{tabular}[c]{@{}c@{}}Process 1, page A\end{tabular} &
  \begin{tabular}[c]{@{}c@{}}Process 2\\ flushes the TLB\end{tabular} &
  \begin{tabular}[c]{@{}c@{}}Process 1, page A, \xmark \textbf{\textcolor{Maroon}{\,no trigger}}\end{tabular} \\ 
\bottomrule
\end{tabular}
\begin{tablenotes}
\centering
    \item \blackcmark:This configuration is enabled for the scenario. -: This configuration is disabled for the scenario. 
\end{tablenotes}
\end{threeparttable}
}
\caption{The XPT prefetcher invalidating results on different scenarios. Note that we build a shared memory region to enable two processes to access the same physical page. $^*$In our experiments, we evaluate various core pairings, \textit{e.g.}, Core-0 and Core-1, and Core-0 and Core-15, to investigate the potential impact of core distance on the results. Our finding shows that we always get similar results, which implies that the XPT prefetcher remains unaffected by core distance.  %
}
\label{tb:invalidate}
\end{table*}

The results of the experiment are presented in Figure~\ref{fig:entry_xpt}. The DRAM access latency increases sharply to approximately 350+ cycles after accessing 256 pages, and then remains stable. \textit{This indicates that the number of entries in the XPT prefetcher is 256.} It is worth noting that the gradual increase in latency with the linear slope between the first page and the $256^{th}$ page is a result of queuing delay.\footnote{We hypothesize that the queuing delay observed is due to the prefetcher, as the latency remains unchanged after priming more than 256 entries. 
However, if the delay is attributed to the memory controller, we expect to see a continuous increase in latency beyond 256 entries.}

In order to accurately determine the set associativity of the XPT, we allocate a 16\,GiB memory pool and generate $N+1$ physical pages with identical bits from the $13^{th}$ bit to the $(13+M)^{th}$ bit. We test with varying values of $N$, specifically $N = 4, 8, 16, 32, 64$, in an attempt to infer the number of cache ways. Additionally, we define $M = 256 / N$, which is used to compute the set number of the corresponding way.

For example, if we guess that the XPT prefetcher uses %
a 4-way set-associative cache, we will prime 5 (4+1) entries to map to the same set, and at least one entry should be evicted, thereby stopping prefetching on that particular page. However, we observed that no eviction occurs even when we increased the guessed number of ways to 128. Based on these findings, \textit{we hypothesize that the XPT operates as a fully-associative cache.}

\begin{figure}[t]
    \centering
    \includegraphics[trim=0 1cm 0 0.75cm,width=1\linewidth]{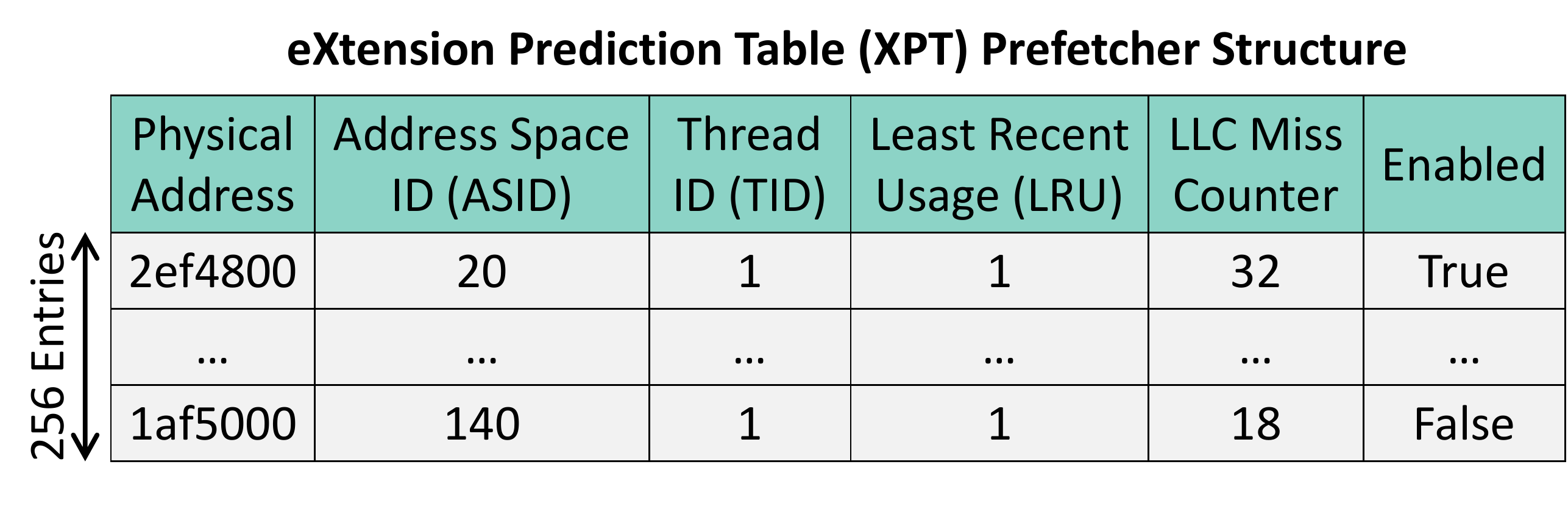}
    \caption{The architecture of the XPT prefetcher}
    \label{fig:arch_xpt}
\end{figure}

\subsection{Key Observations and Leakage Sources}
\label{sec:root-causes}

As we discussed in Section~\ref{sec:xpt-overview}, there are two main root causes leading to 
data leaks from
the XPT prefetecher (XPT evictions and TLB flushes). 
In this section, we will discuss our key observations.
Table~\ref{tb:invalidate} consists of six distinct scenarios.
The first two scenarios involve the creation of two threads running in separate processes, with one of the processes (Process 1) trained on shared page A. Another process (Process 2) is then trained on shared page A or B, and we assess the ability of Process 1 to continue triggering the XPT prefetcher on page A. The results of these two scenarios 
allow us to conclude
that \textit{the XPT is shared among multiple processes}. In addition, the third scenario highlights that the XPT prefetcher is similarly \textit{shared between different threads within the same process space}.

The fourth and fifth scenarios demonstrate different behaviors. 
The fourth scenario suggests that the XPT prefetcher should possess a thread identification tag for controlling its status. As a result, if two threads are executing in the same process space, \textit{i.e.}, they share the same Address Space ID (ASID), and one of the threads accesses a page not present in the XPT, then the well-learned entries associated with that ASID will be invalidated. This raises the question of whether all entries related to the ASID will be invalidated. The fifth scenario was conducted to provide an answer. In this scenario, Thread 1 trains the XPT prefetcher on pages A and B, and then Thread 2 trains the XPT prefetcher on page C. The results showed that only the prefetching of page A was stopped. In another experiment, where Thread 1 trains the XPT prefetcher on pages A, B, and C and then Thread 2 trains on pages D and E, the prefetching of pages A and B was found to have been stopped. These results indicate that a thread will replace entries in the XPT trained by threads with the same ASID, using a Least Recent Usage (LRU) replacement policy.

\begin{center}
\fcolorbox{black}{observationColor}{
\begin{minipage}{0.95\linewidth}
\evictObs{}. When two threads, \texttt{t1} and \texttt{t2}, have the same ASID and \texttt{t1} trains the XPT prefetcher on certain pages. When \texttt{t2} starts a page walk, the oldest page trained by \texttt{t1} will be evicted.
\end{minipage}
}
\end{center}

In order to comprehensively examine the influence of the page management unit, particularly the TLB, on the behavior of the XPT prefetcher, we conducted the sixth scenario. First, Process 1\footnote{It also holds the scenario that two different threads have the same ASID.} trains on a shared page A. Then, Process 2 performs a flush operation on the TLB. Our results indicate that if a TLB flush occurs, the XPT prefetcher cannot be triggered.

\begin{center}
\fcolorbox{black}{observationColor}{
\begin{minipage}{0.95\linewidth}
\flushObs{}. When a page mapping is flushed from the TLB, the corresponding page in the XPT prefetcher will also be flushed.
\end{minipage}
}
\end{center}

\subsection{Summary of the XPT Prefetcher Operations}
The XPT prefetcher, depicted in Figure~\ref{fig:arch_xpt}, operates as follows. Upon accessing a physical page, it checks for a hit in the table. If it is present and the cache miss counter is 32 or higher, then the prefetcher is triggered and a speculative load is issued. When a table miss occurs, the XPT prefetcher checks if the ASID of the physical address is recorded. If present, it then checks the Thread ID (TID). If the TID is found, entries related to this TID are retained, and the new page replaces the oldest entry of other TIDs under the same ASID using the LRU replacement policy. If there are no other TIDs associated with this ASID or if the ASID is not presented in the table, the page is allocated with a new entry with a cache miss counter of 1, unless the prefetcher is full, in which case the oldest entry is replaced using the LRU policy. 

In summary, while the XPT prefetcher's mechanism is promising to be beneficial for memory-intensive applications with irregular memory accesses, our key observations in this section lead to various successful and practical side-channel and covert-channel attacks.

\section{Eviction-based Cross-Core Side-Channel Attacks}
\label{sec:eviction-sca}

In light of our observations from the XPT prefetcher behavior,
we propose a new side-channel attack that exploits these characteristics to leak secret keys from real-world cryptographic applications.
In this section, we focus on \evictObs{} that exploits the contention and eviction policies of the XPT prefetcher.

\subsection{Attacking Square-and-Multiply RSA}
\label{sec:ms_rsa}
We first leverage \evictObs{} to leak the private exponent of the Square-and-Multiply RSA application used in GnuPG 1.4~\cite{gnupg}, which is a real-world application that
has been used as a proof-of-concept by recent work~\cite{guo2022adversarial}. %
As shown in Listing~\ref{listing:ms_rsa}, for every iteration, if the least significant bit (LSB) of the $exp$ is equal to b'1, the base $b$ will be loaded from memory and multiplied with temporary result $r$ (line 3). Otherwise, only a square operation will be executed (line 4). 

\begin{figure}
\centering
    \includegraphics[trim=0 0.5cm 0 0.75cm,width=\linewidth]{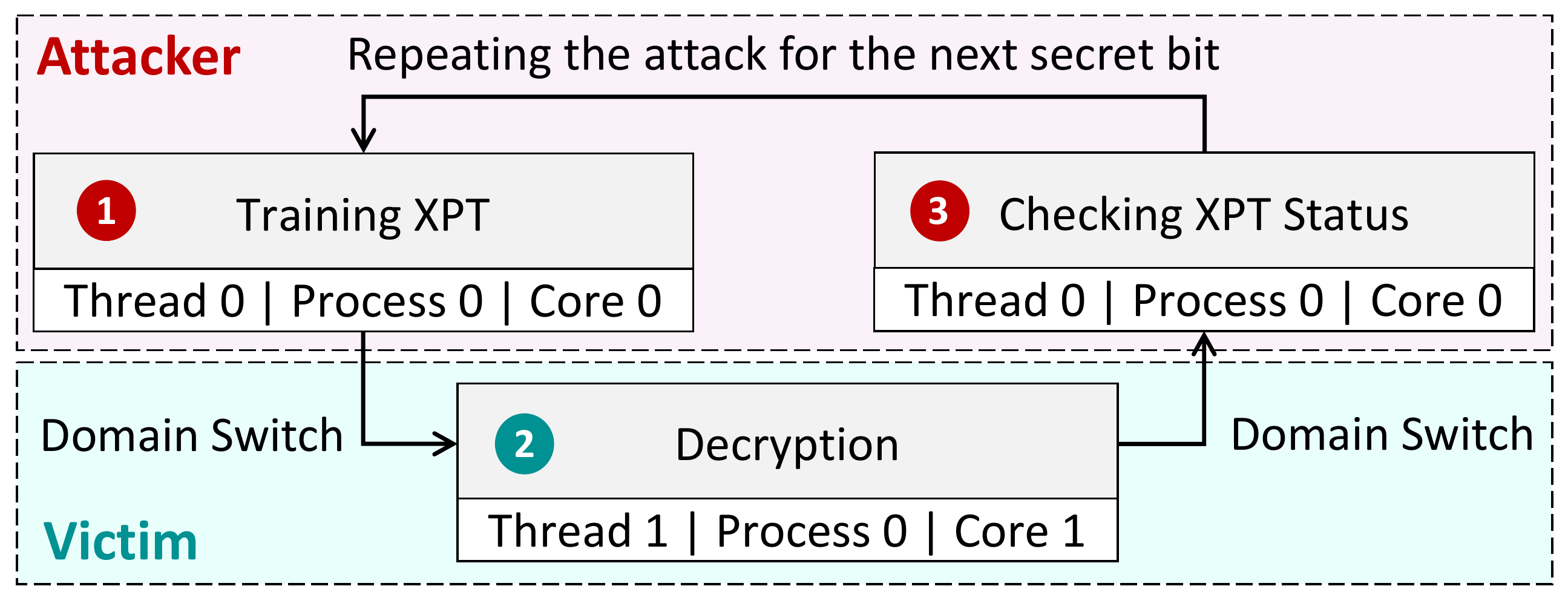}
    \caption{Overview of attacking RSA applications. After the attacker trains the XPT prefetcher, the victim decrypts the message and the attacker then can check the prefetcher status.}
    \label{fig:sync}
\end{figure}

An XPT entry will be evicted if a new thread 
initiates a page walk
(see \evictObs{}). Hence, an attacker can train the XPT prefetcher on a page other than the page storing $b$ and then measure the prefetcher status after each decryption iteration to know if the secret-dependent branch (line 2 in Listing~\ref{listing:ms_rsa}) is executed or not.
Figure~\ref{fig:sync} shows an overview of our attack to reveal the RSA private exponent.
For a successful attack, we need to address two challenges: (1) how to synchronize with the victim thread to guarantee we can measure the XPT prefetcher status at a proper time. (2) How to detect the page used to store $b$ without \texttt{sudo} (as we need to avoid training on this page).

\begin{figure}[t]
\begin{lstlisting}[language=C, caption=Code segment of the Square-and-Multiply RSA in GnuGP., label=listing:ms_rsa]
while (exp) {
  if (exp & 0x01)
    r = (r * b) mod n;//load b from a shared page
  r = (r * r) mod n;
  exp = exp >> 1;
}
\end{lstlisting}
\end{figure}

In this work, we use the \texttt{pthread} and \texttt{semaphore} libraries to solve the fine-grained synchronization challenge with the victim.
More specifically, in our experimental setup, both the attacker and victim threads are created using the \texttt{pthread\_create} function, and the attacker utilizes a \texttt{semaphore} to lock the CPU in order to evict the shared library (which creates a page walking action for the victim to insert the page into the XPT prefetcher when the victim loads $b$) and train the XPT prefetcher. After training the XPT prefetcher, the attacker unlocks the CPU via releasing the \texttt{semaphore}. The victim then locks the CPU and runs the RSA to the attacker's expected point (\textit{e.g.}, a decryption iteration), and then unlocks the CPU which allows the attacker to measure the prefetcher status. 
This method is often used as a simplified synchronization technique in many attacks~\cite{van2020cacheout,gullasch2011cache} and has been demonstrated to not be a critical limitation and can be dealt with by a Linux scheduler attack~\cite{gullasch2011cache}. %

Moreover, concerning the second challenge, the attack can easily uncover the page offset of $b$ by employing the Intel Pin-based technique proposed in Binoculars~\cite{zhao2022binoculars}. Alternatively, the attacker can execute the shared library locally and infer the page mapping via RAMBleed~\cite{kwong2020rambleed}. In this work, we leverage the second technique to uncover page mapping from the user level.
In Section~\ref{sec:rsa-results}, we present our proof-of-concept (PoC) attack results that lead to successful extraction of one byte of the RSA secret key in 2 minutes.

\subsection{Attacking Montgomery Ladder RSA}
\label{sec:constant_rsa}

The Montgomery Ladder RSA~\cite{rsaml}, implemented in popular libraries such as  OpenSSL 1.01~\cite{opensslrsaml} and MbedTLS~\cite{mbed}, is a real-world RSA application that offers resistance against timing side-channel attacks. The key aspect of the algorithm (Listing~\ref{listing:ts_rsa}) is that regardless of the key, the function \texttt{multiply\_and\_add()} is always invoked twice with different inputs. While the execution time of both paths remains constant, 
the order in which these two calls are executed varies based on the secret key. Specifically, if the key is equal to b'1, the values $res1$ are calculated and stored in memory first, followed by the calculation and storage of $res2$. The order is reversed for the secret bit b'0.
As $res1$ and $res2$ are stored in different pages~\cite{zhao2022binoculars}, it is possible to extract the private key by distinguishing the order in which the pages are accessed.

\begin{figure}[t]
\begin{lstlisting}[language=C, caption=Code segment of the Montgomery-Ladder RSA., label=listing:ts_rsa]
while (exp > 0) {
  bool check = exp ? true : false;
  if (check) {
    multiply_and_add(x1, z1, x2, z2, ctx); //res1
    multiple_and_add(x2, z2, x2, z2); //res2
  }
  else {
    multiply_and_add(x2, z2, x1, z1, ctx); //res2
    multiple_and_add(x1, z1, x1, z1); //res1
  }
  exp = exp >> 1;
}
\end{lstlisting}
\end{figure}

To construct the attack, we first determine the page offsets of $res1$ and $res2$ using the technique mentioned in Section~\ref{sec:ms_rsa}. By training the XPT prefetcher on the page storing $res2$ and then checking the prefetcher status on that page after the first \texttt{multiply\_and\_add()} function is complete, we can determine which page has been accessed by the victim. When the secret bit is b'1, a page walk is initiated to the page containing $res1$, resulting in the eviction of the trained entry from the prefetcher. This eviction causes the attacker to observe a normal LLC miss when accessing uncached data on the trained page. In contrast, when the secret bit is b'0, the page holding $res2$ is accessed first, and the well-trained entry is not evicted, which results in an optimized LLC miss.
Similar to our Square-and-Multiply RSA attack, the main challenge is the synchronization with the victim between the two \texttt{multiply\_and\_add()} invocations to measure the XPT prefetcher status. 
Synchronization at function call-level granularity is relatively straightforward (\textit{e.g.}, by manipulating the \texttt{semaphore} at the beginning and end of the \texttt{multiply\_and\_add()}).
Section~\ref{sec:rsa-results} discusses the experimental results of our successful attack to leak one of byte of the Montgomery-Ladder RSA secret key.

\begin{center}
\fcolorbox{black}{usecaseColor}{
\begin{minipage}{0.95\linewidth}
\evictscaUsecase{}. 
We exploit \evictObs{} to reconstruct the secret-dependent behaviors in various implementations of RSA engine and build a side-channel attack.
\end{minipage}
}
\end{center}

\section{TLB-Flush-based Cross-Core Side-Channel Attacks}
\label{sec:tlb_sca}

\begin{figure}[tb!]
    \centering
    \includegraphics[trim=0 0 0 0cm,width=\linewidth]{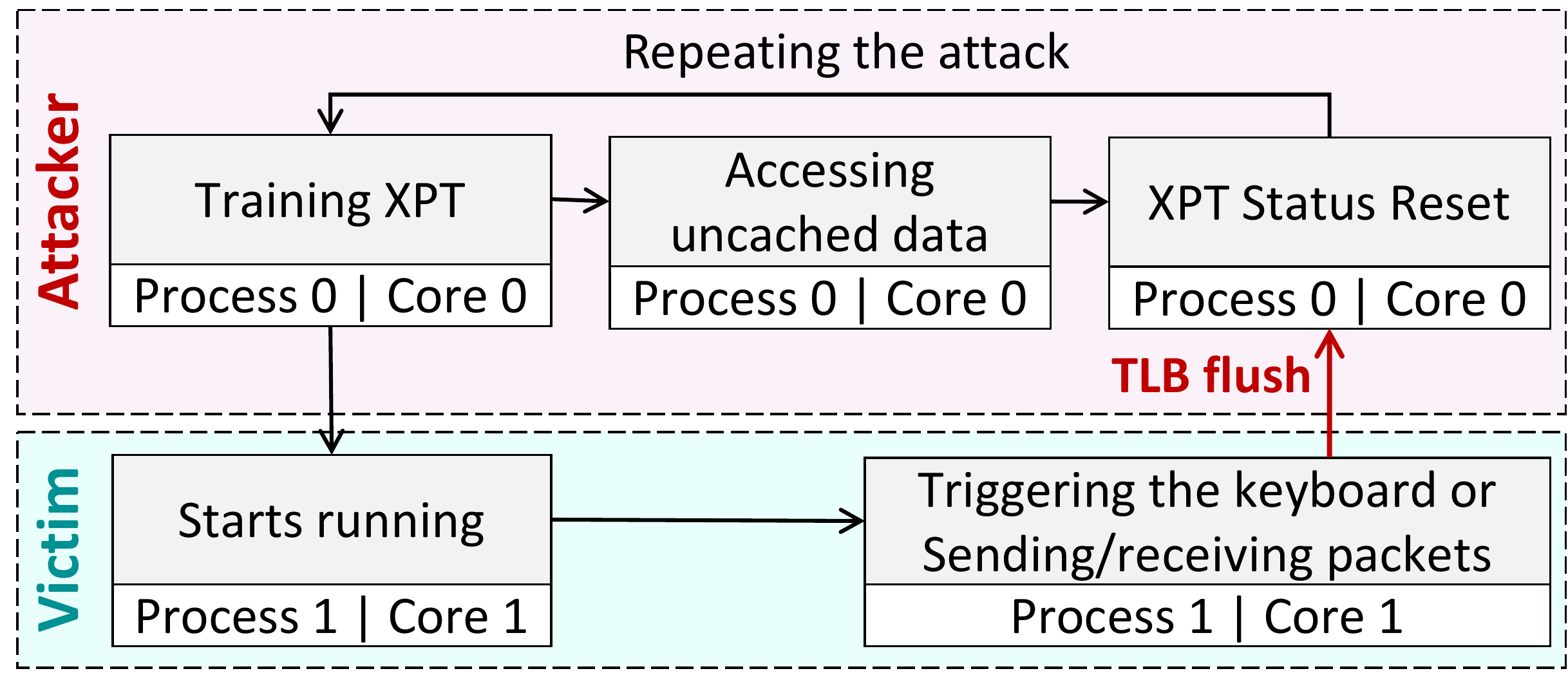}
    \caption{The \name{} flow for TLB-flush-based side-channel attacks.}
    \label{fig:flush_sidechannel_flow}
\end{figure}

Our side-channel attacks presented in this section focus on scenarios in which the victim's sensitive behavior results in TLB flushes (exploiting \flushObs{}). In particular, we have observed that certain events related to drivers (\textit{e.g.}, keystrokes, Bluetooth connections, network packet transmissions via a network card, etc.) can trigger IPIs on Intel processors when running on separate cores with other threads and updating shared memory. This causes a TLB flush and resetting the XPT prefetcher status, resulting in a normal LLC miss latency for uncached data access.
In our attacks,
we run the attack process in parallel with the victim, and the synchronization is done by IPIs signals generated by the victim. Figure~\ref{fig:flush_sidechannel_flow} depicts the \name{} flow for TLB-flush-based side-channel attacks.

\subsection{Monitoring Keyboard Activities}
\label{sec:keyboard-sca}
Our first attack focuses on leaking the precise timing of keystrokes, \textit{i.e.}, when the keyboard is activated by the victim. This leakage is very important since it can assist in reconstructing typed words from users~\cite{keystroke, pnetcat}.

We first consider a victim that %
receives user input from the keyboard (\textit{e.g.}, \texttt{getchar}, \texttt{scanf}, etc.), and writes the input into a shared memory region. 
In case of keyboard activation, the XPT prefetcher resets and will 
no longer trigger,
as demonstrated by \flushObs{}. 
More concretely, in the first step, the attacker trains the XPT prefetcher using the first 32 cache lines of the shared page. The victim then runs on a different core, and the attacker repeatedly tests the XPT prefetcher status. If the XPT prefetcher is triggered, the attacker observes an optimized cache miss, which means the keyboard is not activated by the victim, and vice versa. The attacker will re-train the XPT prefetcher once its status is reset. 
We discuss the results of our attack in Section~\ref{sec:keystroke-results}.

\subsection{Monitoring Network Traffic}
\label{sec:network-sca}

Network traffic analysis attacks~\cite{2002Fingerprinting, 2016finger,2018dns} represent a serious threat to online security. %
Although an attacker cannot get the content of the message from the intercepted message, an attacker can determine the location of both sides of the communication, deanonymize communicating parties, and deduce sensitive information by observing the patterns of datagrams (\textit{e.g.}, packet transmission timing interval, number of transmitted packets, etc.).

We observe that whenever a packet is sent or received by the server or client, the XPT prefetcher status is reset, which is caused by TLB flush (\flushObs{}).
Hence, an attacker is capable of tracking the timing of packet transmission via the XPT prefetcher status.

Listing~\ref{listing:tcp} shows the code segment of the TCP client as our victim. The key parts are lines 6 and 8. Line 6 establishes the connection with the client, while line 8 is used to receive packets and write these packets into a shared buffer.
The victim on Core-1 always tries to receive packets. Similar to our keystroke attack, the attacker on Core-0 also repeatedly checks the XPT prefetcher's status. In case of receiving a packet by the victim, the prefetcher resets and the attacker infers the timing if the prefetcher does not trigger. The attacker will re-train the XPT prefetcher for following rounds of detection.

\begin{center}
\fcolorbox{black}{usecaseColor}{
\begin{minipage}{0.95\linewidth}
\flushscaUsecase{}. 
We exploit \flushObs{} to monitor victim activities that involve flushing the TLB based on a specific and sensitive activity (\textit{e.g.}, keystrokes, network packet transmission, etc.) and construct a side-channel attack.
\end{minipage}
}
\end{center}

\begin{figure}[t]
\begin{lstlisting}[language=C, caption=Code segment of the victim TCP client. Line 6 establishes the connection with the client and line 8 receives the packets., label=listing:tcp]
void victim_client() {
    int sd = socket(AF_INET,SOCK_STREAM,0);
    struct sockaddr_in serveraddr,clientaddr;
    set(serveraddr); //set network protocol
    socklen_t len = sizeof(clientaddr);
    int acceptfd = accept(sd,(struct sockaddr *)&
                             clientaddr,&len);
    int recvbytes = recv(acceptfd,share_buf,
                         sizeof(share_buf),0);
}
\end{lstlisting}
\end{figure}

\section{Cross-Core Covert-Channel Attacks}
\label{sec:covert-channel-attacks}

\textbf{Attack assumptions}. 
Similar to previous covert-channel attacks~\cite{advprefetch, guo2022leaky, host}, the sender and receiver share data via shared pages with each other.
In addition,
the sender and receiver should agree on predefined protocols, including synchronization, encoding, and error correction up-front. %

\subsection{Flush-based Cross-Core Covert-Channel Attack}

\begin{figure}[tb!]
   \centering
   \includegraphics[trim=0 0.4cm 0 1cm,width=0.92\linewidth]{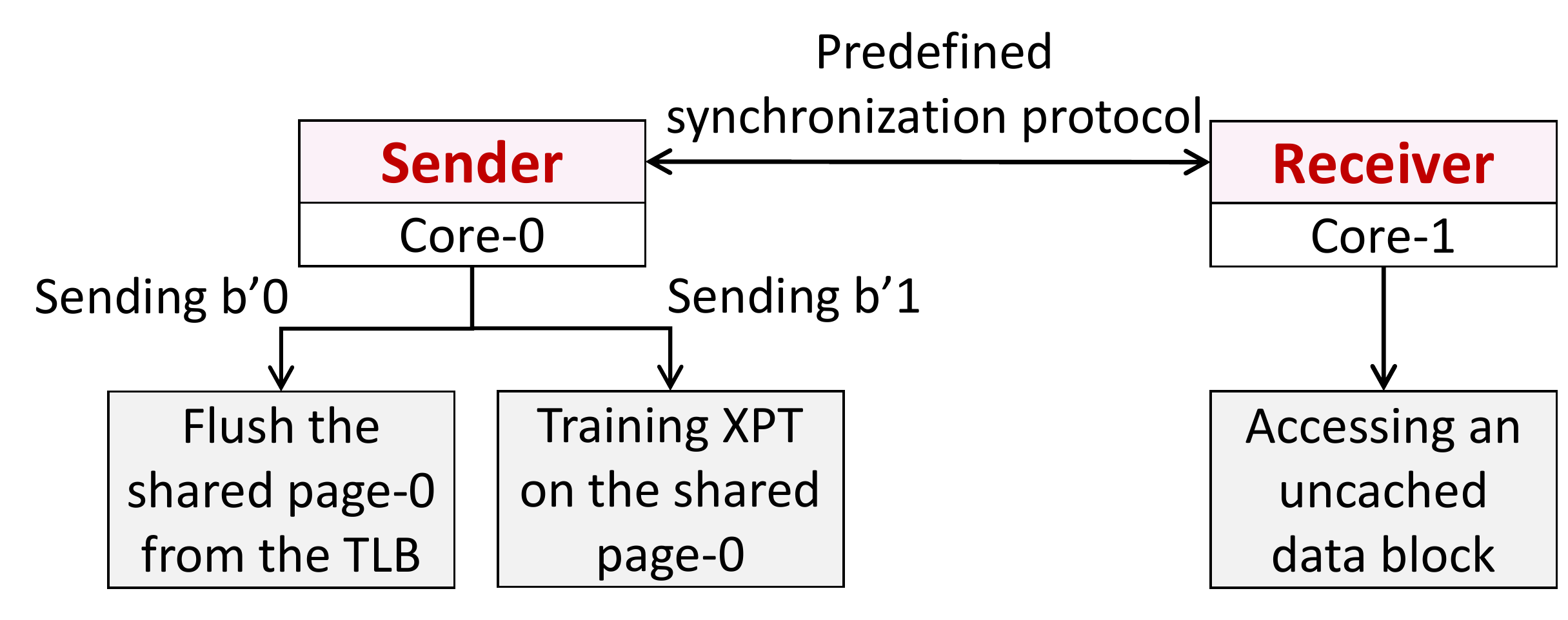}
   \caption{Attack flow of flush-based covert-channel.}
   \label{fig:covert_channel_flush}
\end{figure}

Figure~\ref{fig:covert_channel_flush} depicts our flush-based covert-channel communication flow. If the sender wants to transmit b'1, it first trains the XPT on the first half page (first 32 cache lines) on Core-0. The training function is the same as the \texttt{train()} used in Listing~\ref{listing:xpt_trigger}. If the sender still wants to transmit b'1 in the next round, it does not need to re-train the XPT as the XPT prefetcher status will be reset only if pages are flushed from TLB or evicted from the prefetcher. Thus, the sender needs to reset the XPT prefetcher status only if he/she wants to transmit b'0 in the next round. After the sender trained/reset the XPT prefetcher, the receiver, running on Core-1, randomly accesses a cache line on the last half page of the shared page to observe if the cache miss is an optimized LLC miss 
or a normal LLC miss (XPT is not triggered). An optimized cache miss indicates that the sender did not reset the XPT, sending b'1, and vice versa.

\begin{table}[t]
\centering
\scalebox{0.8}{
\begin{tabular}{c||c|c|c|c}
\toprule
\multirow{2}{*}{Secret Pattern} & \multicolumn{2}{c|}{Flush-based} & \multicolumn{2}{c}{Eviction-based} \\
\cline{2-5}
& Throughput & Error Rate & Throughput & Error Rate \\
\midrule
 $(111...111)_2$ & 1.7\,MiB/s &  0\% & 1.7\,MiB/s &  0\% \\ 
\hline
 $(101...010)_2$ &  21\,KiB/s & 0.5\% & 49\,KiB/s & 8\% \\
\bottomrule
\end{tabular}}
\caption{Covert-channel attack experiments using Intel XPT prefetcher.}
\label{tb:covert-channel-results}
\end{table}

We conducted experiments to evaluate the performance of our proposed covert-channel attack, by measuring the channel throughput and error rate under different secrets. Our results show that in the best case where the sender always sends the same bit (\textit{i.e.}, b'1111...1111, see the first row of Table~\ref{tb:covert-channel-results}), the throughput can achieve up to 1.7\,MiB/s. Even in the worst case where the binary pattern alternates between 1s and 0s, the throughput is still able to achieve 21\,KiB/s. 
In both cases, the error rate of the covert-channel is less than 1\%. Our results demonstrate the feasibility and robustness of our proposed covert-channel exploiting \flushObs{}. 

\subsection{Eviction-based Covert-Channel Attack}

\begin{figure}[tb!]
   \centering
   \includegraphics[trim=0 0.5cm 0 1cm,width=0.92\linewidth]{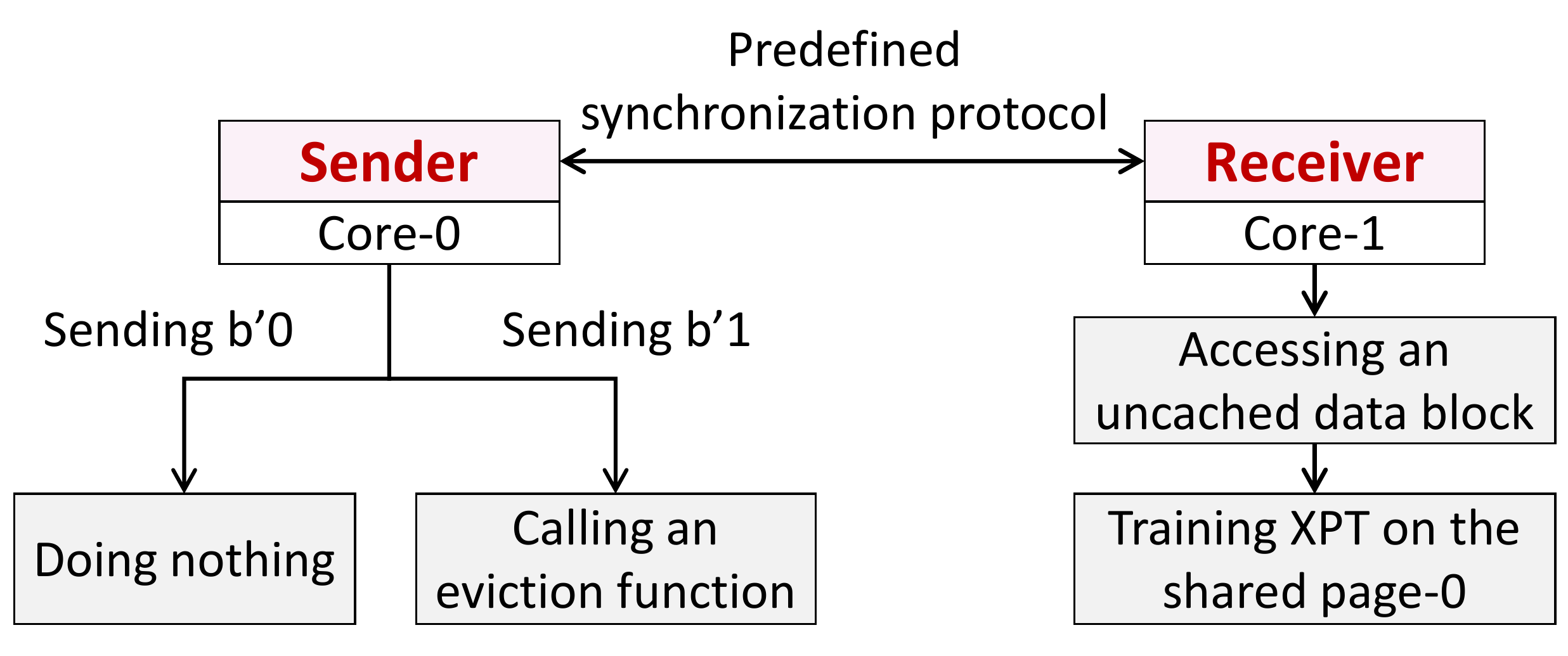}
   \caption{Attack flow of eviction-based covert-channel.}
   \label{fig:covert_channel_evict}
\end{figure}

Figure~\ref{fig:covert_channel_evict} illustrates the flow of the eviction-based covert-channel attack. In contrast to the flush-based covert-channel attack, where the sender trains the XPT prefetcher first, the receiver trains the XPT prefetcher on a specific page. The receiver then continually tests whether it can still trigger the prefetcher. To transmit b'1, the sender invokes an eviction function that trains an unmapped shared page and thus evicts the page from the XPT prefetcher that was trained by the receiver. On the other hand, the sender does not need to execute any function to transmit b'0, other than synchronizing with the receiver.

In the best case scenario (\textit{i.e.}, b'11...11), the achieved throughput 
is equivalent to that of a flush-based covert-channel, which can reach up to 1.7\,MiB/s (see Table~\ref{tb:covert-channel-results}). Importantly, in the worst-case scenario (\textit{i.e.}, b'101...010), the attained throughput is 49\,KiB/s, which is twice the rate of the flush-based covert-channel. It should be noted that the error rate in this scenario is higher than the flush-based covert-channel, as some unexpected page walks are performed when the receiver thread switches to the attacker thread (we assume that it can be caused by the OS or the page prefetcher introduced by Intel~\cite{npp}) and evicts the well-trained entry.

\section{Experimental Results}
\label{sec:expset}

\textbf{Experimental environment.}
We perform Proof-of-Concept (PoC) side-channel experiments on an Ice Lake machine. The system details are shown in Table~\ref{tb:conf}.  
\begin{table}[t]
\small
\centering
\scalebox{0.9}
{%
\begin{tabular}{c|c}
\toprule
AWS EC2 Instance & m6i.4xlarge          \\ \hline
Processor        & Intel(R) Xeon(R) Platinum 8375C          \\ \hline
Architecture     & Ice Lake (Sunny Cove)             \\ \hline
CPU cores        & 16                   \\ \hline
Last Level Cache & Non-inclusive, 54\,MiB  \\ \hline
Operating System & Ubuntu 20.04         \\ \hline
ASLR/KASLR       & Enabled              \\ \hline
DRAM             & DDR4, 64\,GiB        \\ 
\bottomrule
\end{tabular}}
\caption{Architecture and system configurations.}
\label{tb:conf}
\end{table}

\begin{figure*}[t]
    \centering
    \includegraphics[trim=0 0.3cm 0 0cm,width=\linewidth]{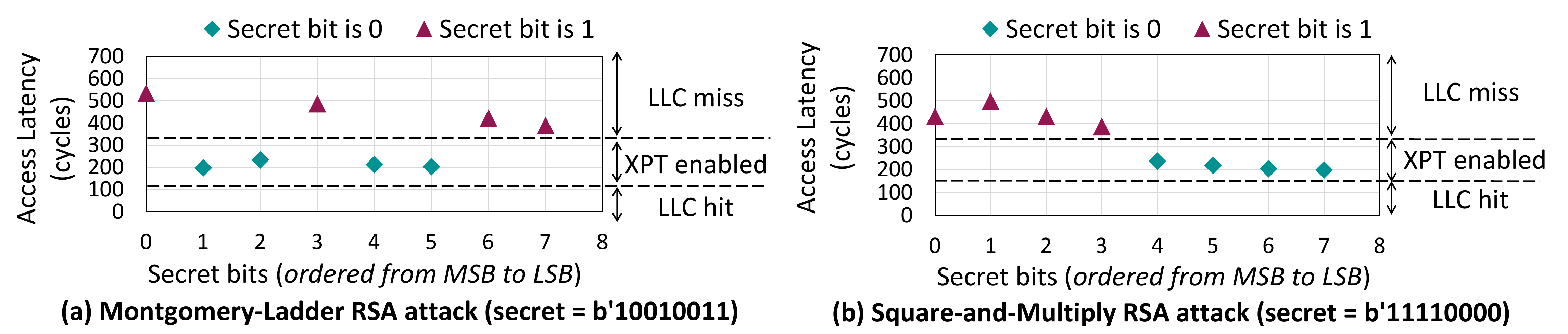}
    \caption{RSA attack results. (a) Montgomery-ladder RSA, and (b) Square-and-Multiply RSA. The x-axis orders the secret bits from most significant bit (MSB) to least significant bit (LSB).
    Note, that in case of b'0 as secret bit, a well-trained entry in XPT is accessed which results to optimized LLC miss.}
    \label{fig:rsa_attack_all}
\end{figure*}

\subsection{RSA Side-Channel Attacks}
\label{sec:rsa-results}

In our PoC, we map the Square-and-Multiply RSA and Montgomery-ladder RSA computations to shared memory and utilize the synchronization and page offset detection technique mentioned in Section~\ref{sec:ms_rsa}.
The Montgomery-Ladder RSA attack result is demonstrated in Figure~\ref{fig:rsa_attack_all}(a). We attempted to leak a 1-byte private exponent (\texttt{0x93}). The results show that we successfully revealed it.
Similar to the Montgomery-ladder RSA implementation, we successfully leak a 1-byte private exponent (\texttt{0xf0}) from the Square-and-Multiply RSA. The experimental result is presented in Figure~\ref{fig:rsa_attack_all}(b). 

For both RSA implementations, we ran 10 iterations to leak one bit of the private exponent, which consumed approximately 2 minutes to leak 1 byte. As every bit's computation is independent in RSA decryption, it is reasonable to estimate that \name{} can break a 2048-bit (256-byte) RSA engine in approximately 512 minutes (8.4 hours), which is a practical time window for an attacker. 

\begin{figure}[t]
    \centering
    \includegraphics[trim=0 0.3cm 0 0.5cm,width=1\linewidth]{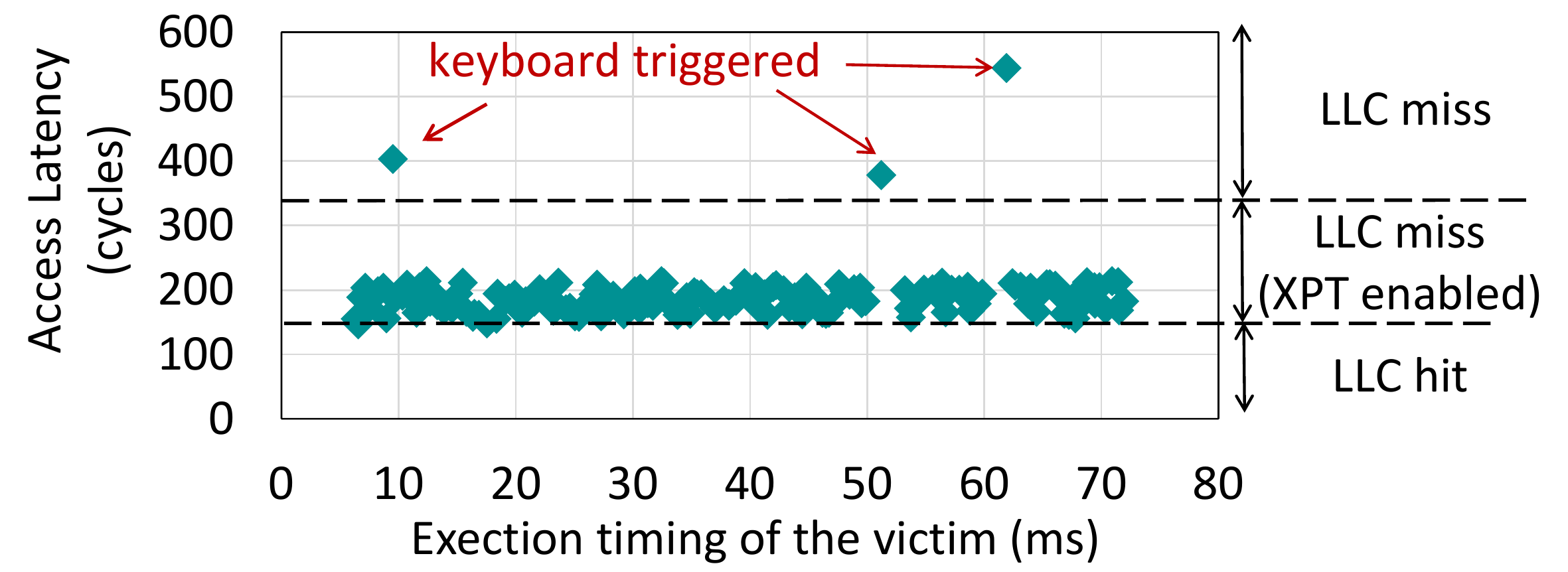}
    \caption{Side-channel attack to monitor user keyboard activities.}
    \label{fig:keyboard-attack}
\end{figure}

\subsection{Keystroke Side-Channel Attack}
\label{sec:keystroke-results}
Figure~\ref{fig:keyboard-attack} presents the results of our keystroke attack. We clearly observe that invoking the keyboard activity function resets the status of the XPT prefetcher. Specifically, the significant difference between the access latency of normal LLC misses and optimized LLC misses allows us to identify the precise timing of each keystroke. 
Table~\ref{tb:keyboard} summarizes the average latency we obtained under different scenarios. The results indicate that without the XPT prefetcher, we could not detect keyboard activity. This finding provides further evidence that the XPT prefetcher plays a crucial role in our attacks.

\subsection{Network Traffic Side-Channel Attack}
\label{sec:network-results}
\begin{figure}[t]
    \centering
    \includegraphics[trim=0 0.3cm 0 0.5cm,width=1\linewidth]{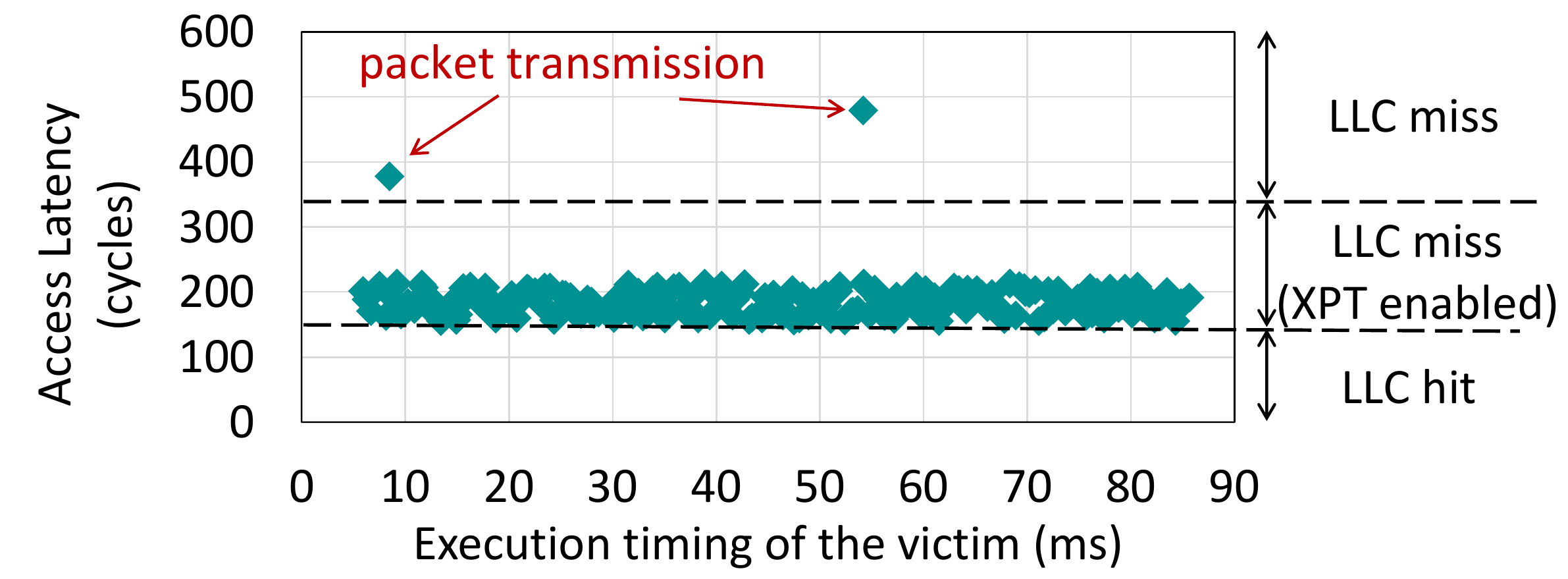}
    \caption{Side-channel attack to monitor user network traffic.}
    \label{fig:network-attack}
\end{figure}

Our PoC involves a server-client setup where the server communicates with the victim client (see Listing~\ref{listing:tcp}) via port 8888. The server randomly generates some packets and then sends them to the client. The outcome of this attack is shown in Figure~\ref{fig:network-attack}, which clearly shows the timing of packet reception by the victim.
Our results demonstrate that by observing the XPT prefetcher status, we can detect network traffic patterns that reveal information about the victim's behavior.

\begin{table}[tb!]
\centering
\scalebox{0.85}{
\begin{tabular}{c||c|c|c}
\toprule
\multirow{3}{*}{Scenario} 
 & \multicolumn{2}{c|}{Memory Access Latency (cycles)} & \multirow{3}{*}{Distinguishable?}\\
\cline{2-3}
 & keyboard & keyboard & \\ 
& activated & inactivated & \\
\midrule
XPT    & 378                & 174       & \cmark{}           \\ \hline
no-XPT & 388                & 383       & \xmark{}           \\ 
\bottomrule
\end{tabular}}
\caption{Average memory access latency whether a keystroke has occurred in different scenarios: XPT enabled or disabled.}
\label{tb:keyboard}
\end{table}

\textbf{Resolution Analysis.} In the real world, the frequency of network packet transmission can be very high, 
and we need to ensure that training the XPT prefetcher is shorter than packet transmission interval\footnote{Keystrokes require milliseconds to seconds to complete, which trivially the XPT prefetcher can be trained in this interval.}.
To investigate this, we sampled IP packets from our network interface card (NIC) and realized that the timing interval between sending two packets are consistently around  $26,000$ nanoseconds (ns).
Training the XPT requires generating 32 LLC misses, which typically take 400 cycles per miss. On our experiment platform, 1 cycle takes about 0.3\,ns. Therefore, in the worst case where there is no memory-level parallelism (MLP) and out-of-order (OoO) execution, we require $32\times400\times0.3$\,ns (or $3,840$\,ns) to train the XPT prefetcher. Considering the MLP and OoO execution, in our real-world measurements, it typically takes around $3,000$\,ns. The significant gap between the packet transmission interval and the XPT prefetcher training time demonstrates that \name{}'s resolution is high enough to track network packets in real-world applications.

\section{Potential Mitigation}
\label{sec:mitigation}

While disabling the prefetcher blocks the leakages of the XPT prefetcher, it may introduce unexpected performance overheads for memory-intensive applications with highly irregular memory accesses. 
A more efficient mitigation is to partition the prefetcher. To achieve leakage-free partitioning, the system needs to partition the XPT prefetcher based on the ASID and the TID tags that already exist in the XPT structure. This ensures to block of information leakage across different cores and different threads on the same core.
Note, that clearing the XPT prefetcher at context switches will not be effective for cross-core and hyperthreading setups.

\section{Related Work}
\label{sec:related-work}

\begin{table*}[t]
\centering
\scalebox{0.8}{
\begin{threeparttable}[b]
\begin{tabular}{c|c|c|c|c|c|c|c|c|c|c}
\toprule
  \multirow{3}{*}{\begin{tabular}[c]{@{}c@{}}Trigger\\  From\end{tabular}}  &
  \multirow{3}{*}{Paper} &
  \multirow{3}{*}{\begin{tabular}[c]{@{}c@{}}Cross\\  Core?\end{tabular}} &
  \multirow{3}{*}{\begin{tabular}[c]{@{}c@{}}Side-\\  Channel?~\tnote{$*$}\end{tabular}} &
  \multirow{3}{*}{\begin{tabular}[c]{@{}c@{}}Hardware\\Leakage\\  Source\end{tabular}} &
  \multicolumn{6}{c}{Requirements}\\
  \cline{6-11}
   & & & & &\rotatebox[origin=c]{90}{\parbox{2cm}{\centering Cache\\  Primitives}} &
  \rotatebox[origin=c]{90}{\parbox{2cm}{\centering Shared\\  Memory}} &
  \rotatebox[origin=c]{90}{\parbox{2cm}{\centering Address of \\ Instruction}} &
  \rotatebox[origin=c]{90}{\parbox{2cm}{\centering Address \\ of Data~\tnote{$\ddagger$}}} &
  \rotatebox[origin=c]{90}{\parbox{2cm}{\centering Speculative\\ Execution}} &
  \rotatebox[origin=c]{90}{\parbox{2cm}{\centering Algorithm\\ Specific}} \\ 
  \midrule
\multirow{3}{*}{\rotatebox[origin=c]{90}{\parbox{1cm}{\centering Sofware}}} & Prefetch Attack~\cite{gruss2016prefetch, lipp2022amd}     &  \textcolor{Maroon}{\xmark} & \cmark & L1/L2/LLC & \fullcirc  & \fullcirc  & \emptycirc & \halfcirc & \emptycirc & \emptycirc\\ 
\cline{2-11}
& Leaky Way~\cite{guo2022leaky}  & \cmark & \textcolor{Maroon}{\xmark}  & LLC &\emptycirc & \emptycirc & \emptycirc & \fullcirc & \emptycirc &\emptycirc  \\ 
\cline{2-11}
& Adversarial Prefetch~\cite{guo2022adversarial} & \cmark & \cmark & LLC & \fullcirc & \fullcirc & \emptycirc &  \fullcirc & \emptycirc & \emptycirc \\ 
\hline

\multirow{5}{*}{\rotatebox[origin=c]{90}{\parbox{3cm}{\centering Hardware}}} & Fetching Tale~\cite{host} & \textcolor{Maroon}{\xmark} & \textcolor{Maroon}{\xmark} & \begin{tabular}[c]{@{}c@{}}IP-Stride Prefetcher\end{tabular} & \emptycirc & \emptycirc & \fullcirc & \emptycirc & \emptycirc & \emptycirc\\   

\cline{2-11}
& Unveiling Prefetcher~\cite{ccs} & \textcolor{Maroon}{\xmark} & \cmark & \begin{tabular}[c]{@{}c@{}}IP-Stride Prefetcher +\\  L1/L2/LLC\end{tabular} & \fullcirc & \fullcirc & \emptycirc & \fullcirc & \emptycirc & \fullcirc\\

\cline{2-11}

& Augury~\cite{vicarte2022augury}  & \textcolor{Maroon}{\xmark} & \cmark &  \begin{tabular}[c]{@{}c@{}}Pointer-Chasing Prefetcher +\\ L1/L2/LLC\end{tabular} & \fullcirc & \emptycirc & \fullcirc & \emptycirc & \fullcirc & \emptycirc \\ 

\cline{2-11}

& AfterImage~\cite{chen2023afterimage}  & \textcolor{Maroon}{\xmark} & \cmark  & IP-Stride Prefetcher & \emptycirc & \emptycirc & \fullcirc & \emptycirc & \emptycirc & \emptycirc\\

\cline{2-11} \noalign{\vspace{0.01cm}}

\rowcolor{JungleGreen!10}%
& \name{} (\textit{this work}) & \cmark & \cmark & XPT Prefetcher& \emptycirc & \fullcirc & \emptycirc & \halfcirc & \emptycirc & \emptycirc  \\ 
\bottomrule
\end{tabular}
\begin{tablenotes}
    \item [$*$] \cmark~: can be used as both side-channel and covert-channel attacks, \textcolor{Maroon}{\xmark}: can only be used as a covert-channel attack
    \item [$\ddagger$] \emptycirc~: no need for the address of the victim's data, \halfcirc~: need the page-grained address, \fullcirc~: need the cache-line-grained address
\end{tablenotes}
\end{threeparttable}
}%
\caption{Summary of prefetcher/prefetch-based attacks. %
}
\label{tb:related}
\end{table*}

\textbf{Cache timing attacks}.
Cache timing side-channels exploit the timing differences between cache hits and cache misses, enabling attackers to infer the victim's memory activities. There are two primary types of cache timing side-channels: Prime+Probe based~\cite{osvik2006cache,percival2005cache,disselkoen2017prime+,papp} and Flush+Reload based~\cite{yarom2014flush+,gruss2016flush+}. For the Prime+Probe type, the attacker primes the cache with their data and then probes it to observe the victim's access time changes. Flush+Reload involves the attacker flushing a cache line and then waiting for the victim to reload it, thereby providing insight into the victim's memory access patterns. Cache timing side-channels are often referred as cache primitives since they serve as a building block for other hardware side-channels~\cite{vicarte2022augury,ccs,van2020cacheout}.

\textbf{Prefetcher attacks}.
Side-channel attacks introduced by prefetching have received extensive attention in recent years. From the software perspective, modern processors often provide specific prefetch instructions that programmers can use to improve performance. Some attacks~\cite{gruss2016prefetch, lipp2022amd} aim to bypass Supervisor Mode Access Prevention (SMAP) and KASLR on Intel and AMD processors. They exploit the timing of \texttt{PREFETCH} instructions to leak the translation level of the virtual address and infer the physical mapping. Leaky Way~\cite{guo2022leaky} exploits \texttt{PREFETCHNTA} in Intel processors to change the cache status and build a covert-channel attack through conflicted cache ways. Another instruction, \texttt{PREFETCHW}, can leak cache coherent status and allow cross-core attacks~\cite{guo2022adversarial}.

On the contrary, hardware prefetchers in real processors cannot be explicitly controlled and normally require more understanding through reverse-engineering. Augury~\cite{vicarte2022augury} investigated the data memory-dependent prefetcher in the Apple M1 processor to perform out-of-bounds reads. AfterImage~\cite{chen2023afterimage} studied Intel IP-based stride prefetcher that enables tracking load instructions of the victim. Other works \cite{shin2018unveiling,host} are either algorithm dependent or can only be used as a covert-channel. However, all existing prefetcher side-channels are inner-core and they are more or less dependent on the cache hierarchy, while \name{} can launch cross-core side-channel and covert-channel attacks that do not rely on the cache system. Table~\ref{tb:related} shows a summary of existing prefetcher/prefetch-based side-channel and covert-channel attacks.

\section{Conclusion}
\label{sec:conclusion}
In this work, we uncover details of the XPT prefetcher in Intel processors, which can be intentionally trained and triggered across different cores within the same processor. Capitalizing on these features, we introduce a novel attack, named \name{}, capable of leaking victims' page activities. \name{} is a cross-core attack, independent of cache primitives as the foundation of many hardware attacks. To achieve this, we conducted an in-depth study of the XPT prefetcher, revealing undocumented details. We demonstrate that we can
extract secret keys in real-world Square-and-Multiply and Montgomery-Ladder RSA applications. Furthermore, we apply \name{} to effectively leak the victim's driver-related events. Additionally, we showcase the applicability of \name{} as cross-core covert-channel attacks, achieving high throughput and low error rates when transmitting secrets.
Finally, we conclude that the processors require either disable the XPT prefetcher or perform a thread-based partitioning to mitigate the \name{} attacks.

\bibliographystyle{ACM-Reference-Format}
\bibliography{ref}

\end{document}